\documentclass[ms,nonblindrev]{informs3_homepage}
\OneAndAHalfSpacedXI 

\usepackage{mystylefile}

\numberwithin{equation}{section}

\usepackage{natbib}
 \bibpunct[, ]{(}{)}{,}{a}{}{,}%
 \def\bibfont{\small}%
\TheoremsNumberedThrough     
\ECRepeatTheorems

\EquationsNumberedBySection 

\MANUSCRIPTNO{MS-0001-1922.65}

\makeatletter
\let\normalsize\relax
\let\@currsize\normalsize
\makeatother

\definecolor{stanfordred}{rgb}{0.55, 0.08, 0.08}
\definecolor{uwpurple}{RGB}{247, 151, 29}

\begin{document}



\RUNTITLE{Learning to Recommend Using Non-Uniform Data}

\TITLE{Learning to Recommend Using Non-Uniform Data}


\ARTICLEAUTHORS{%
   \AUTHOR{Wanning Chen}
	\AFF{Graduate School of Business, Stanford University, \EMAIL{wanningc@stanford.edu}}
	\AUTHOR{Mohsen Bayati}
	\AFF{
		Graduate School of Business, Stanford University, \EMAIL{bayati@stanford.edu}}
	} 

\ABSTRACT{Learning user preferences for products based on their past purchases or reviews is at the cornerstone of modern recommendation engines. One complication in this learning task is that some users are more likely to purchase products or review them, and some products are more likely to be purchased or reviewed by the users. This non-uniform pattern degrades the power of many existing recommendation algorithms, as they assume that the observed data are sampled uniformly at random among user-product pairs. In addition, existing literature on modeling non-uniformity either assume user interests are independent of the products, or lack theoretical understanding. In this paper, we first model the user-product preferences as a partially observed matrix with non-uniform observation pattern. Next, building on the literature about low-rank matrix estimation, we introduce a new weighted trace-norm penalized regression to predict unobserved values of the matrix. We then prove an upper bound for the prediction error of our proposed approach. Our upper bound is a function of a number of parameters that are based on a certain \emph{weight matrix} that depends on the joint distribution of users and products. Utilizing this observation, we introduce a new optimization problem to select a weight matrix that minimizes the upper bound on the prediction error. The final product is a new estimator, NU-Recommend, that outperforms existing methods in both synthetic and real datasets. Our approach aims at accurate predictions for all users while prioritizing fairness. To achieve this, we employ a bias-variance tradeoff mechanism that ensures good overall prediction performance without compromising the predictive accuracy for less active users.}%


\KEYWORDS{personalization, recommender systems, collaborative filtering, matrix completion, high-dimensional learning}

\maketitle
\section{Introduction}

Recommendations are now ubiquitous and powerful in various settings such as streaming services, e-commerce and social networks. As shown in \cite{McKinsey}, 35 percent of what consumers purchase on Amazon and 75 percent of what they watch on Netflix come from product recommendations. Personalized recommendation, an effort to suggest to users different products so as to tailor to their different needs or tastes, has boosted views, number of sold items and overall sales for companies \citep{mobilegame}. It also vastly increases click-through and conversion rates than the untargeted content such as banner advertisements and top-seller lists \citep{amazon}.

Collaborative-filtering-based algorithms are the backbones of many personalized recommender system to estimate user preferences well. These predictive algorithms analyze information from existing data such as ratings, purchase history and click-through rates. They predict user's preference by leveraging information not just from this single user's past experience, but also from that of other users, especially those who have similar interest to this single user. 

Matrix completion methods are a popular family of collaborative filtering algorithms, but the majority of them have theoretical guarantees under the unrealistic assumption that the data are observed uniformly at random. They approach the recommendation task as one of recovering a preference matrix from its partially observed entries. The rows of the matrix correspond to users, the columns correspond to items and the entries contain user preferences. Informally speaking, such uniform observation assumption entails that all users will equally likely rate, purchase, or click on items and the items will equally likely be rated, purchased, or clicked on by the users. But in reality, such assumption may not hold in most scenarios. For example, in the rating case, some users are more active than others and some items are rated by many people while others are rarely rated. 
 
\cite{NIPS2010_4102}, \cite{foygel} and \cite{mnar} demonstrated empirically that if we go beyond the assumption that entries are revealed with the same probability independently, we can substantially improve the accuracy of inferring user preferences. It is mentioned in \cite{NIPS2010_4102} that the uniformly at random assumption is not just a deficiency of the proof techniques. It also leads to a significant deterioration in prediction quality and an increase in the sample complexity. Thus, these aspects point towards a need to include the non-uniform aspect of the sampling scheme when learning user preferences, so as to improve prediction accuracy. 

We consider an example of a popular matrix completion approach, known as trace-norm penalized regression, and show how we effectively include the non-uniform sampling process in the prediction task, building on the existing literature. The objective of the trace-norm penalized regression is to predict observed entries close to the observed values and at the same time penalize a convex surrogate of the `matrix rank', i.e., the trace-norm of the matrix. \cite{NIPS2010_4102} suggested to correct the penalty term by marginal distributions of the rows and the columns. However, as pointed out by \cite{foygel}, such `margin-based' weighted correction becomes suboptimal when the sampling scheme is not a product distribution, which occurs when user preferences is not independent of the products. Indeed, our analysis shows that the margin weighting strategy can be further improved. 

Before we detail our contribution, we want to point out that our algorithm can be adopted to improve personalized treatment recommendations in the healthcare industry as well. According to \cite{healthcare}, developing a machine learning-based automated model that can provide laboratory test recommendation has a great potential to support current clinical practice. Works such as \cite{clinicnet} and \cite{TRD} applied machine learning to improve lab test results inference. Matrix-completion-based collaborative filtering can handle shared information across different patients and different lab tests in a smart manner. The automation resembles traditional practice of physicians prescribing lab tests to a patient based on similar past cases and suggests to a physician which medical tests to order for a patient based on existing lab test results in the electronic health records (EHR). These treatment recommendations are personalized so as to meet different idiosyncrasies in patient health conditions. With the help from recommender systems, a physician can better navigate through and efficiently use huge amount of information from hundreds and thousands of patients' medical records. This process, if implemented correctly, can enhance physicians' ability to include all potentially relevant and not-so-obvious medical tests for a patient and hence reduce the risk of under-testing; in addition, it can save a healthcare organization the unnecessary expense on irrelevant test orders. It can further alleviate physician stress and burnout caused by extended hours working with complex information from EHR \citep{burnout}. 

In the healthcare setting, medical test measurements are not taken uniformly at random, while the patients are not equally likely to be tested. We will demonstrate how our method can be used to learn lab test results under the non-uniform sampling pattern among patient-medical test pairs, without assuming that the patient health conditions and medical tests are independent. Since our method is non-context based, we do not need any patient feature information or lab test feature information. Thus, if the hospital needs to hand the data to a third party to operate the recommendation system, the patient information can be protected.

\subsection{Our approach and contributions}
The algorithm we propose is a novel theory-driven algorithm that learns user preferences under a more general non-uniform sampling scheme, which allows user preferences to depend on the products. We provide theoretical guarantees for our approach and also show that it outperforms existing benchmarks on synthetic and real data.

\paragraph{A unified weighted framework and its error bound.} The highlight of our approach is that we provide an upper error bound on how accurate the estimation of the underlying ground truth preference matrix is, for a unified weighted trace-norm penalized estimator that we come up with under non-uniform sampling scheme. This general estimator encompasses two well-studied objectives as special cases: standard trace-norm penalized regression \citep{recht} and marginal weighted-trace-norm penalized regression \citep{NIPS2010_4102}. Unlike them, we regularize a weighted version of our parameters and at the same time not restrict the weights to be of rank one. Our more general bound reduces to the existing known results for these two objectives when the special weights are plugged into the general formulation. Our work differs significantly from previous studies that use such upper bounds for qualitative assessment of the algorithm. Instead, we utilize the upper bound to come up with essential inputs of our algorithm.

\paragraph{Our Algorithm.} By minimizing the upper error bound we derive, we come up with a theory-driven and effective weighting strategy that leverages the distribution of observed user-product pairs. We call this algorithm NU-Recommend (`NU' stands for non-uniform). Our weighting strategy benefits from successfully capturing the interaction between the sampling matrix and the underlying preference matrix.

\paragraph{Empirical results.} We complement our theory-driven algorithm NU-Recommend with empirical experiments on both synthetic and real data. Our benchmarks include standard unweighted strategy, the margin weighting strategy, an inverse propensity weighting strategy, and a universal singular value thresholding strategy modified to account for non-uniformity. Two real-world datasets are deployed: user rating data and medical test results data. The former is used to show our algorithm NU-Recommend's superior performance on imputing user preferences based on observed rating data, and the latter is used to show that NU-Recommend can help with lab test recommendations in the healthcare setting.

\paragraph{Fairness restoration.} We also conduct a straightforward analysis to show that, by our way of incorporating the non-uniform sampling pattern, fairness can be restored in terms of estimating less observed users and products more accurately comparing to the aforementioned benchmarks. This is achieved by a more equitable bias-variance trade-off, due to our specially designed weighting strategy.

\subsection{Other related work}
Our work can be viewed as an instance of designing matrix-completion-based collaborative filtering method to learn user preferences, in order to enhance personalized recommender systems (see \cite{recbook} for a comprehensive overview of personalized recommendation algorithms). 

There are more direct measurements of the effectiveness of recommender systems, but to improve the quality of user preference prediction is believed to be at the foundation of enhanced recommendations. Indeed, the accuracy of predicting user preferences is very much valued such that the well-known Netflix Data Competition, launched in 2006, awarded the winner one million dollars for their smallest error on estimating how someone is going to rate a movie. Learning user preferences is also the core input for many other operational and marketing activities besides product recommendations, including personalizing search results, designing loyalty programs and delivering one-to-one marketing. We defer to  \cite{farias2019learning} and references therein for further discussion on this. 

The adoption of collaborative filtering to learn users' preferences dates back to the early stage of Amazon's effort to build recommendation engines \citep{amazon}, and it is still highly sought-after \citep{wu2022survey}. There is a substantial amount of literature for matrix completion models for collaborative-filtering-based recommender systems \citep{mcsurvey}. Some popular approaches include matrix factorization \citep{srebromf, salakhutdinovmf, keshavan1, keshavan}, rank minimization \citep{bunea, klopprank} and the aforementioned trace-norm penalized regression \citep{recht, tao, plan}. In fact, matrix factorization is one of the two algorithms used by the winner of Netflix Prize, and which Netflix later chose to put into production \citep{netflixneverimplement}. As mentioned before, one distinct difference between our method and these canonical works is the following: we consider a more general sampling distribution, whereas most of them assumed the restrictive uniform or margin-based sampling scheme while designing their algorithm as well as providing theoretical guarantees.

There are a few existing works on matrix completion for non-uniform sampling pattern. Besides the aforementioned margin weighting strategy proposed by \cite{NIPS2010_4102}, \cite{foygel} proposed a smoothing variant of it. However, their method requires knowledge of the rank of the matrix, which may be unknown in practice. \cite{mnar} channeled the inverse propensity weighting (IPW) technique from the causal inference literature to modify the trace-norm penalized regression. Such weighting strategy addresses the selection bias by constructing a weighted regression based on propensity scores of the observed entries. \cite{mnar} showed effectiveness of weighting with the non-uniform propensity score matrix through empirical experiments, but their approach lacks theoretical guarantees. \cite{bhattacharya2022matrix} proposed a modified universal singular value thresholding (USVT) method accustomed to the data missing pattern. Their method is computationally fast but suffers from poor statistical accuracy in empirical performance. \cite{WMC-IEEE} proposed a rank-$r$ projection of the margin-based weighted correction, analogous to nuclear norm penalization of the margin-based weighted correction in \cite{NIPS2010_4102}, mixed with the IPW strategy in \cite{mnar}. Again, their weighting strategy focuses on rank-1 weight matrix whereas we allow a more general weight matrix. \cite{chen2015completing} designed an active sampling strategy to collect data based on the leverage scores which describe the local mass concentration of a matrix, in order to efficiently recover a noiseless matrix, whereas we work with a given dataset.

To provide properties for the general weighted-trace-norm penalized regression that we propose, we deploy techniques from high-dimensional statistics literature. A key challenge in our proof is that we assume the weighting strategy to be arbitrary. \cite{sahand} studied theoretically the marginal weighting strategy proposed by \cite{NIPS2010_4102} via assuming the sampling scheme to be a product distribution. In addition, they directly take the sampling matrix to be the weighting matrix, which is encompassed by our more general setting. Furthermore, our proof techniques are distinct from theirs since they used a more involved $\epsilon$-net argument while we build on ideas developed in \cite{klopp} and \cite{nima}. We note that \cite{nima} considered the trace-regression problem under a general sampling scheme. However, for the matrix completion problem, their approach is equivalent to the uniform sampling scheme.

For the non-uniform setting, we will be working under the regime of missing completely at random (MCAR), meaning that the missingness pattern is independent of the values of the underlying ground truth preference matrix. We acknowledge that most observational datasets require less restrictive missing data mechanisms such as missing at random (MAR) and missing not at random (MNAR), but addressing the non-uniform missing patterns is an important building block for future work.

To account for the dependency of the observation pattern on the underlying matrix values or the non-uniform noises, one needs to use causal inference methods which rely on an estimation method that addresses non-uniform sampling.  We refer the readers to \cite{kartik} for a discussion on how to model the missing mechanism more generally, and to \cite{mnarmc} for their approach on blending trace-norm penalized regression with MNAR and the reference therein.

\subsection{Organization} 
The rest of this paper is organized as follows. We describe the problem formulation in Section \ref{probform}. We present the weighted-trace-norm penalized estimator and our main result on the estimator's theoretical performance in Section \ref{sec:alg}, as well as the design of our new algorithm `NU-Recommend'. In Section \ref{theory}, we provide a proof sketch for our main theorem. Finally, empirical results on simulated data as well as the MovieLens data and medical test data are presented in Section \ref{sec:emp}, followed by a discussion on restoring fairness with our approach due to bias-variance tradeoff. Proofs are relegated to the appendices.

\section{Model and Problem}
\label{probform}
We use bold capital letters (e.g. $\bB$) for matrices and non-bold capital letters for vectors (e.g., $V$). For a positive integer $m$, we denote the set of integers $\{1,\ldots,m\}$ by $[m]$. For a matrix $\bB$, $\bB_{jk}$ refers to its entry $(j,k)$. 
 
We will use rating data as a running example. We encode the ground truth user preference data in a $d_r\times d_c$ matrix called $\bB^*$, and the row indices $j \in[d_r]$ of matrix $\bB^*$ correspond to users, the column indices $k\in[d_c]$ correspond to items.  That is, the $(j,k)$-th element $\bB^*_{jk}$ is the rating of user $j$ towards item $k$. Suppose we have $n$ number of observations. For $i\in [n]$, let $(j_i, k_i)$ be the $i$-th observed sample and its observed value is $y_i$. 
 
 In the rest of this section, we will specify the data generating model, the sampling model and also some additional structures we impose on $\bB^*$ to make it recoverable.

\subsection{Data Generating Model}
We assume that $y_i$ is the noisy realization of entry $(j_i, k_i)$ such that $y_i = \bB^*_{j_ik_i} + \epsilon_i$, where $\epsilon_i$ is a noise term. Let $Y = [y_1,y_2,\dots,y_n]^\top$ be the $n$ by $1$ vector of observed values and $E =[\epsilon_1, \epsilon_2,\dots, \epsilon_n]^\top$ be the vector of independent mean zero noise random variables with variance at most $\sigma^2$. For any positive integer $m$, $e_1(m), e_2(m), \dots, e_m(m)$ denote the standard basis vectors for $\mathbb{R}^m$. Define the design matrix $\bX_i := e_{j_i}(d_r)e_{k_i}^{\top}(d_c)$, that is, $\bX_i$ is everywhere zero except a single one at entry $(j_i, k_i)$. Then $\bB^*_{j_ik_i} = \langle \bB^*, \bX_{i} \rangle$, where $\langle\cdot, \cdot\rangle$ denotes the trace inner product of two matrices such that 
\[
\langle \bB_1, \bB_2\rangle:= \Tr(\bB_1\bB_2^{\top})\,.
\]
To make the notation more succinct, let us define a sampling operator which takes in a matrix $\bB$ and outputs a $n$-dimensional vector (i.e. $\mathfrak{X}:\mathbb{R}^{d_r\times d_c}\rightarrow \mathbb{R}^n$). The operator takes the following form: $$[\mathfrak{X}(\bB)]_i := \langle \bB, \bX_{i} \rangle.$$ Elements of this vector are the entries of $\bB$ at $n$ observed locations. Our data generating model can then be written as 
\[
Y = \mathfrak{X}(\bB^*) + E.
\]
This model is called the trace regression model \citep{trevor}. Our estimation problem is that of estimating $\bB^*$, having observed $Y$ and the design matrices $\bX_i$, $i\in [n]$. There are also generalized trace regression models to deal with non-linear associations, which conceptually can be applied here as well. We restrict our study to the linear setting to avoid further complication of the notation.

\subsection{General Sampling Model}
To model the sampling distribution, let $\bP^*$ encode the rating probabilities where $\bP^*_{jk}$ denotes the probability that the $(j,k)$-th entry is observed, i.e. probability that user $j$ will rate item $k$, and $\sum_{j\in[d_r]}\sum_{k\in[d_c]}\bP^*_{jk} = 1$ (up to normalization). Then the observation pattern could be viewed as a ``noisy realization'' of this underlying sampling matrix. The uniform sampling assumption made in the past literature means that $\bP^*_{jk}$ are all equal for $j\in [d_r]$ and $k\in [d_c]$. This is equivalent to say that each user will equally likely rate different items and each item will be rated equally likely by different users, which usually is not the case. 

Instead, we want to consider a general sampling matrix such that entries in $\bP^*$ are not necessarily equal to each other. Equivalently, this is saying that each $\bX_i$ is sampled independently from a distribution $\Pi$ on the set of canonical basis matrices for $\mathbb{R}^{d_r\times d_c}$: $\Big\{e_j(d_r)e_k^{\top}(d_c), 1\leq j\leq d_r, 1\leq k\leq d_c\Big\},$ such that 
\[
\bP^*_{jk} = \mathbb{P}\Big(\bX = e_j(d_r)e_k^{\top}(d_c)\Big) = \Pi_{jk}\,.
\]

We further assume that each element is sampled with positive probability and the probabilities are bounded.
\begin{assumption}
\label{assump_p_min_max}
There exist a positive constant $p_{\min}$, that may depend on $d_r$ and $d_c$, such that 
\[
p_{\min}\leq \bP_{jk}.
\]
\end{assumption}

Note that $p_{\min}\leq \frac{1}{d_rd_c}$ and the equality is achieved in the case of uniform sampling scheme.
By Assumption \ref{assump_p_min_max}, we only focus on cases that $p_{\min}$ is not too small (for example, it can be a constant multiple of $\frac{1}{d_rd_c}$). To understand the validity of this assumption, companies would often treat the long tails and light tails algorithmically differently. In our case, when confronting very small sampling rate, we can work with a (denser) submatrix whose entry-wise sampling probabilities satisfy Assumption \ref{assump_p_min_max}, and then deploy different technologies such as prior business benchmarks for the rest of the matrix with a small $p_{\min}$.

A less general weighted sampling model considered in literature is to only assume a marginal probability on observing each user and a marginal probability on observing each item, denoted by $R\in \mathbb{R}^{d_r}$ and $C\in \mathbb{R}^{d_c}$ respectively. They are also called row marginal probability and column marginal probability. To be more specific, $R_j$ is the probability that user $j$ is sampled and $C_k$ is the probability that item $k$ is sampled, and $\sum_{j\in [d_r]}R_j = 1$, $\sum_{k\in [d_c]}C_k = 1$. Then the probability of observing entry $(j,k)$ is $R_jC_k$. Note that, if the underlying entry-wise sampling distribution $\bP^*$ is $RC^{\top}$, i.e. a product distribution, then the marginal model is the same as the general sampling model we have proposed, since $\bP^*_{jk} = R_jC_k$, $R_j = \sum_{k\in[d_c]}\bP^*_{jk}$ and $C_k = \sum_{j\in [d_r]}\bP^*_{jk}$. However, this sampling model is limiting $\bP^*$ as a product distribution, i.e. a rank-1 sampling matrix, which is a rather restrictive assumption. Consider the following example when $\bP^*$ is not a product distribution: $\bP^* = \begin{bmatrix}0.2 & 0.3 \\0.3 & 0.2\end{bmatrix}$. $\bP^*$ is a rank-2 matrix and the entries are not uniformly sampled. The row marginal probability $R = [ 0.5, 0.5]$ and column marginal probability $C = [0.5, 0.5]$. According to the marginal sampling model, the probability of observing each entry is 0.25, thus each entry is assumed to be uniformly sampled, which is not true. As such, assuming an entry-wise probability covers more broad cases than only assuming the  marginal probability.

\subsection{Low-rankness}
\paragraph{Low-rankness of the preference matrix.} 

In many instances, the matrix we wish to recover is known to be low-rank. We recall that, by definition, a matrix of $d_r\times d_c$ has rank $r$ if its rows or columns span a $r$-dimensional space.

In reality, users typically rate only very few items so that there are very few scattered observed entries of this ground truth matrix. That is, the number of observations $n$ is comparably much smaller than the total number of entries $d_rd_c$, which is also the number of parameters needed to be recovered. Therefore, without any additional structure, the estimation problem is \emph{high-dimensional}; it is an under-determined system of equations with many solutions which would overfit the observed data and the noise. 

Hence, a small $r$, i.e. low-rankness, is very desirable in such high-dimensional estimation problems. With such a low-rank structure, we only have roughly $r(d_r+d_c)$ free parameters to learn, which translates to a sample size of this order (up to logarithmic factors) to control for the recovery error. As such, we have turned the problem into a low-rank matrix completion problem for which many methods have been designed. 

Intuitively, the low-rank structure entails that only a few factors contribute to a user's tastes or preferences. This can be explained better in terms of the latent factor model. The preference matrix $\bB^*$ can be written as product of two matrices: user latent factor matrix $\bU\in\mathbb{R}^{d_r\times r}$ and item latent factor matrix $\bV\in\mathbb{R}^{d_c\times r}$ such that $\bB^* = \bU\bV^{\top}$. In another word, the rating of user $j$ towards item $k$, $\bB^*_{jk}$, equals the dot product between $U_j$ and $V_k$, where $U_j$ is the $j$-th row of $\bU$ and $V_k$ is the $k$-th row of $\bV$. We only need $r$ number of factors to explain each user and each item, and $r$ is much small than the number of users and items, $d_r$ and $d_c$.

\begin{figure}
    \centering
    \includegraphics[scale = 0.55]{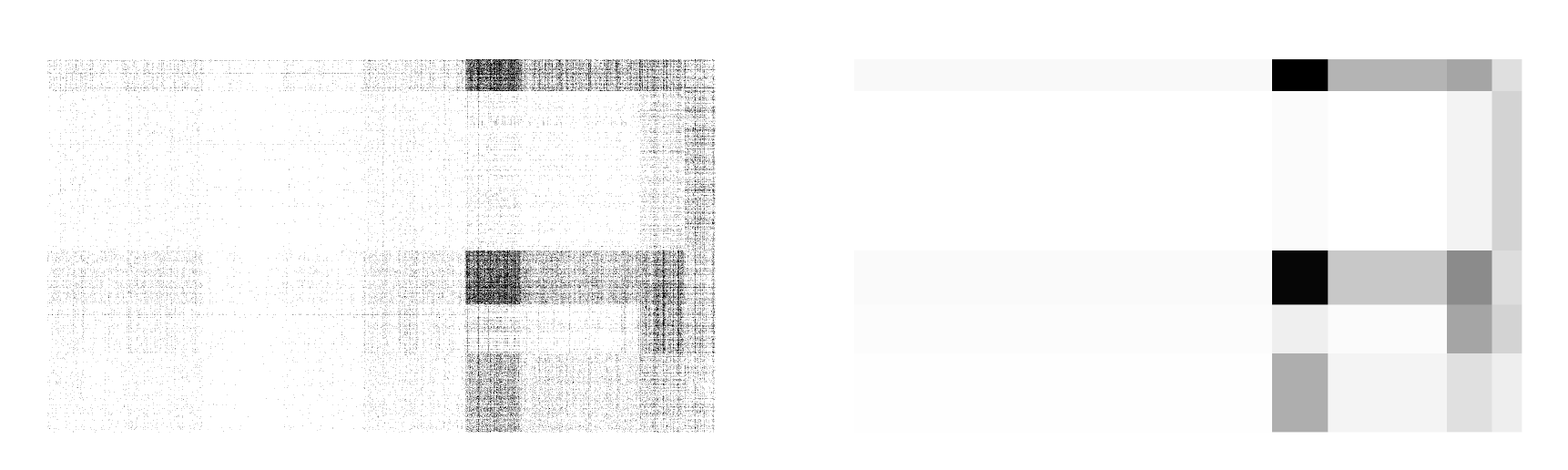}
    \caption{Missingness matrix for MovieLens-100k dataset. Black indicates an entry being observed. On the left is the missingness matrix and on the right is its corresponding block structure identified using spectral biclustering mentioned in \cite{biclustering}; rows and columns have been rearranged based on the biclustering result.}
    \label{fig:my_label}
\end{figure}

\paragraph{Low-rankness of the sampling matrix.} Sampling-based observation pattern with a low rank structure has been studied recently. In \cite{mnar}, they presented a missingness matrix of the MovieLens-100k dataset (see Figure \ref{fig:my_label}) where the rows are users, columns are movies, and a black dot indicates an observed entry. Such missingness pattern with few block structures suggests that the observations are generated from a low rank sampling matrix $\bP^*$. This structure can be very useful when estimating the sampling distribution.

\subsection{Spikiness} 
It is also known that low-rank structure alone is not enough to recover the matrix. Consider a ground truth user preference matrix that is everywhere else equal to zero except for a single entry that is equal to one. Then it is impossible to recover such spiky matrix unless the entry which contains the single one is observed. Therefore, the recoverability of a matrix also depends on its spikiness. \cite{nima} defined a more natural and less restrictive way to formalize notion of spikiness by the following form:
\[
\text{spikiness of an arbitrary matrix }\bB:=\frac{\mathfrak{N}(\bB)}{\|\bB\|_F},
\]
where $\mathfrak{N}(\bB)$ is some norm of $\bB$ that depends on the distribution of the design matrices. We will use this notion of spikiness and provide specific form of $\mathfrak{N}(\cdot)$ in our analysis.

\subsection{Our Problem}
Our task of estimating $\bB^*$ based on the noisy observations $Y$ is combined with the assumptions that the sampling model $\bP^*$ is a general one, both $\bB^*$ and $\bP^*$ are low rank and that the ground truth matrix $\bB^*$ satisfies certain ``spikiness" condition. The goal is to construct an estimator $\widehat \bB$ to minimize some loss function with respect to $\bB^*$. A common loss function is the Frobenius error:  $$\|\bB^* - \widehat\bB\|_F$$ where $\|\bB\|_F:= \sqrt{\sum_{(j,k)\in [d_r]\times[d_c]}\bB_{jk}^2}$ for a matrix $\bB\in \mathbb{R}^{d_r\times d_c}$. However, this error does not reflect the non-uniform sampling nature of the problem since it is putting equal weights to the errors of all entries regardless of each entry's sampling probability. 

Instead, we study the convergence properties of the $L^2(\Pi)$ error: $$\|\bB^* - \widehat\bB\|_{L^2(\Pi)}$$ where $\|\bB\|_{L^2(\Pi)}:=\sqrt{\mathbb{E}[\langle \bB,\bX\rangle^2]}$, in which the expectation is taken with respect to a probability measure $\Pi$ on $\mathbb{R}^{d_r\times d_c}$ from which we sample the random matrix $\bX$. This loss function will weigh the error of each entry according to the sampling distribution, and it reflects the test RMSE more readily. 

Empirically, for the synthetic data, we will present both the relative frobenius error $\frac{\|\bB^* - \widehat\bB\|_F}{\|\bB^*\|_F}$ and the relative $L^2(\Pi)$ error $\frac{\|\bB^* - \widehat\bB\|_{L^2(\Pi)}}{\|\bB^*\|_{L^2(\Pi)}}$ for performance comparison; for the real data, since the ground truth matrix $\bB^*$ is not known, we will present the test RMSE
\[\sqrt{\frac{1}{n_{te}}\sum_{(j,k)\in \Omega_{te}}(\bB^*_{jk} - \widehat\bB_{jk})^2}\,,
\]
where $n_{te}$ is the number of test data and $\Omega_{te}$ denotes the set of (row, column) indices of test entries.

\section{NU-Recommend Algorithm}
\label{sec:alg}

The weighted matrix completion algorithm NU-Recommend we propose builds on the aforementioned trace-norm penalized regression. The classic trace-norm penalized regression objective solves the following convex program: 
\begin{align}
 \min_{\bB\in \mathbb{R}^{d_r\times d_c}}\frac{1}{n}\|Y - \opr(\bB)\|_2^2 + \lambda \|\bB\|_*.
 \label{regular-trace-norm}
\end{align}
The trace norm (also called the nuclear norm) of a matrix $\bB\in \mathbb{R}^{d_r\times d_c}$ is defined to be the sum of its singular values, i.e., $\|\bB\|_*:=\sum_{i\in[r]} \bD_{ii}$ where $\bD$ is a $r\times r$ diagonal matrix in the singular value decomposition (SVD) of $\bB = \bU\bD\bV^{\top}$ and $r$ denotes the rank of $\bB$. In this formulation, the potential  to incorporate information about the sampling distribution is missing. To address this shortcoming, \cite{NIPS2010_4102} proposed the following margin-weighted trace-norm penalized regression:
\begin{align}
 \min_{\bB\in \mathbb{R}^{d_r\times d_c}}\frac{1}{n}\|Y - \opr(\bB)\|_2^2 + \lambda \|\diag{R}\bB\diag{C}\|_*,
 \label{margin-trace-norm}
\end{align}
where $\diag{R}$ is a diagonal matrix with the row marginal probability $R$ lying on the diagonal and $\diag{C}$ is a diagonal matrix with the column marginal probability $C$ lying on the diagonal. The term inside the nuclear norm $\diag{R}\bB\diag{C}$ is equivalent to weighting the matrix $\bB$ by a matrix $\bW$ element-wise where $\bW = RC^{\top}$, i.e. $\bW \circ \bB$. Here, $\circ$ denotes the element-wise product (Hadamard product) between two matrices. \cite{sahand} analyzed such rank-1 weighting correction under the scenario where the sampling matrix is also the rank-1 matrix $RC^{\top}$. 

Our goal is to generalize the formulation under general sampling schemes to the following convex program that we call the weighted-trace-norm penalized regression, where $\bW$ is not specified at this point yet:
\begin{align}
 \min_{\bB\in \mathbb{R}^{d_r\times d_c}}\frac{1}{n}\|Y - \opr(\bB)\|_2^2 + \lambda \|\bW\circ\bB\|_*.
 \label{weighted-trace-norm}
\end{align}
Note that if we take $\bW$ to be a matrix of all ones, then this formulation reduces to (\ref{regular-trace-norm}). So our formulation covers both the regular trace-norm penalized regression and the margin-weighted trace-norm penalized regression.

Under a general sampling scheme $\bP^*$, do we simply match the weight matrix $\bW$ to be the sampling matrix $\bP^*$, or is there some other way to design a weight matrix which delivers better performance? The answer becomes obvious through the recovery guarantee of this formulation we present next. But clearly, we want the weight matrix $\bW$ to inherit some information about the sampling matrix $\bP^*$ and to be not too far off from $\bP^*$. To measure the closeness between $\bW$ and $\bP^*$, we introduce the following scalar term $l$:
\begin{definition}
\label{def_l}
We define $l$, which is a function of a weight matrix $\bW$ and the sampling matrix $\bP^*$, to be the smallest constant that satisfies
\[
\frac{1}{l}\leq \frac{\bP^*_{jk}}{\bW_{jk}}\leq l.
\]
\end{definition}
With this definition, making $\bW$ close to $\bP^*$ translates into picking a weight matrix $\bW$ so that $l$ is not too large.

\subsection{Main Result: Recovery Guarantee of Weighted Objective}
We assume that $d_r = d_c = d$ to simplify the notation, but the statement also holds for $d_r\neq d_c$. Let $n^* = d^2\|\sqrt{\bW}\circ \bB\|_{\infty}$ and $\tr$ denote the rank of $\bN^* = \sqrt{\bW}\circ \bB^*$. Our main result states that the weighted objective produces an estimator that is consistent and that the error bound is an increasing function in $l$, $n^*$ and $\tilde r$.
\begin{theorem}
Let $\sigma^2$ be the noise variance. We have:
\[
\|\widehat\bB-\bB^*\|^2_{L^2(\Pi)}\leq C_1 (\sigma^2\vee {n^*}^2)\frac{d\rho l^4\tilde r}{np_{\min}},
\]
with probability at least $1-\exp(-C_2\rho)$ for some $\rho\geq \log d$ and constants $C_1, C_2$ and $C_3$, whenever $n\geq C_3(\log\frac{l^2}{p_{\min}d})^2\frac{\log d}{p_{\min}d}$.
\label{matcompthm}
\end{theorem}

When $\bW$ and $\bP$ are both equal to the matrix with all entries equal to $1/d^2$, i.e. $p_{\min} = \frac{1}{d^2}$ and $l = 1$, then this result reduces to the bound analyzed by \cite{nima} for the regular (non-weighted) trace-norm penalized regression under uniform sampling scheme. In other words, our bound is more general and reduces to the same rate for unweighted trace-norm penalized regression under the uniform sampling assumption. To obtain a proper weight matrix $\bW$ under the non-uniform reality, we optimize over the derived error bound.

\subsection{Weight Matrix Construction}
\label{weight-mat-constr}
As shown in the upper bound, the error indeed increases in $l$. Although taking $\bW = \bP^*$ makes $l=1$, the smallest possible, there are additional terms that depend on $\bW$. One such term is $\tilde r$, which is also a function of $\bW$; so is the term $n^*$ that controls the spikiness of the weighted ground truth matrix. Motivated by this, we want to minimize these three terms, $l$, $\tilde r$ and $n^*$, simultaneously. From the low rank matrix completion literature, minimizing $\tilde r$ can be well approximated by minimizing its convex surrogate $\|\sqrt{\bW}\circ \bB^*\|_*$. While minimizing this nuclear norm, we add bounds on $l$ and $n^*$ to keep them small. Now by introducing the variable $\bQ = \sqrt{\bW}$, we propose the following optimization problem where the square root is taken element-wise, and $l$ and $\gamma$ are hyper-parameters:
\begin{align}
& \min_{\bQ} \|\bQ \circ \bB^*\|_* \\ \nonumber
\text{ subject to } & \frac{1}{l}\leq \frac{\bQ}{\sqrt{\bP^*}}\leq l \text{ and } \|\bQ \circ \bB^*\|_{\infty} \leq \gamma.
\label{weight-construction}
\end{align}
Once a solution $\bQ$ to the above minimization is found, one can take $\bW$ to be the normalized version of $\bQ^2$, where $\bQ^2$ denotes elementwise square of $\bQ$, i.e. $$\bW = \frac{\bQ^2}{\sum_{jk}(\bQ_{jk}^2)}.$$

Of course, in reality, we do not have access to the ground truth $\bB^*$ and $\bP^*$. Instead, we will use two estimates: a raw estimator of the ground truth matrix $\widehat\bB$ (e.g. which can be obtained via other aforementioned existing methods) and an estimator of the sampling matrix $\widehat\bP$, and plug them in the following program:
\begin{align}
& \min_{\bQ} \|\bQ \circ \widehat\bB\|_* \\ \nonumber
\text{ subject to } & \frac{1}{l}\leq \frac{\bQ}{\sqrt{\widehat\bP}}\leq l \text{ and } \|\bQ \circ \widehat\bB\|_{\infty} \leq \gamma.
\label{weight-construction}
\end{align}
Our NU-Recommend Algorithm first constructs a matrix $\bW$ with Program (\ref{weight-construction}) and then plugs it in the weighted objective (\ref{weighted-trace-norm}) to obtain a refined estimator $\widehat\bB$. An alternative approach is to feed the refined estimator into (\ref{weight-construction}) to repeat the process, but it does not yield further improvement in our numerical simulations.

\subsection{Estimation of the Sampling Matrix}
\label{sec:samp-est}

As we have specified in our sampling model, the sampling matrix $\bP^*$ does not necessarily have equal entries and it has a low rank structure. We can estimate it through the design matrices $\bX_i$, $i \in [n]$, by letting $\bM\in \mathbb{R}^{d_r\times d_c}$ denote the observation matrix which records how many times each entry is observed. That is, $\bM_{jk} = \sum_{i\in [n]} \mathbbm{1}\Big\{\bX_i = e_j(d_r)e_k^{\top}(d_c)\Big\}$. We introduce two ways to estimate the sampling matrix: rank-1 estimation and Poisson matrix recovery.

\subsubsection{Rank-1 estimation}
\paragraph{} As mentioned in \cite{NIPS2010_4102}, if we assume that $\bP^*$ is rank-1, then we can first estimate the marginal probability $\widehat R$ and $\widehat C$ by the following formula:
\[
\widehat R_j = \sum_{k\in [d_c]}\bM_{jk}/n \quad \text{and} \quad \widehat C_k = \sum_{j\in [d_r]}\bM_{jk}/n.
\]
Then the rank-1 estimation of $\bP^*$ is $\widehat\bP = \widehat R\widehat C^{\top}$. However, this estimation may not be accurate enough as it simplifies the rank of $\bP^*$ to 1. Next, we present a more sophisticated method that does not restrict the rank of $\bP^*$ to 1.
\subsubsection{Estimation through Poisson matrix recovery}
\paragraph{Poisson model.} We model the entries in the observation matrix with Poisson model where these entries are Poisson counts of the product of the corresponding entry in the sampling matrix and the number of samples, i.e., $\bM_{jk}\sim \text{Poisson}(n\bP^*_{jk})$. By definition of Poisson distribution, this is to say that each entry $(j,k)$ in expectation has $n\bP^*_{jk}$ number of observations when the sample size is $n$ and $$\mathbb{P}(\bM_{jk} = m) = \frac{(n\bP^*_{jk})^me^{-(n\bP^*_{jk})}}{m!}.$$ 
\paragraph{Poisson matrix recovery.} Note that we are trying to recover the low-rank sampling matrix based on the realizations of the entries. Thus, this is a low-rank matrix recovery problem. We adapt a Poisson matrix recovery method called PMLSVT introduced in \cite{poisson}, which is a variant of the proximal gradient descent method applied to the following objective: 
\begin{align*}
    \min_{\bP} -\sum_{(j,k)\in \Omega} (\bM_{jk}\log \bP_{jk} - \bP_{jk}) + \lambda \|\bP\|_*,
\end{align*}
where $\Omega$ is the set of indices of observed entries. Note that this objective is essentially a nuclear norm penalized maximum log-likelihood estimator for the Poisson model. Algorithm \ref{PMLSVT} presents the pseudo-code for PMLSVT. 
\begin{algorithm}[h]
\caption{Low-rank sampling matrix estimation using PMLSVT}
\label{PMLSVT}
\begin{algorithmic}[1]
\State Initialize: The maximum number of iterations $K$, parameters $\eta$ and $t$. $\bX \leftarrow \bM$.
\For{$\lambda = \lambda_1, \lambda_2, \dots, \lambda_L$}
\For{$k = 1, 2, \dots, K$}
\State $\bC\leftarrow \bX - (1/t)\nabla f(\bX)$
\State $\bC = \bU\bSigma\bV^T\{\text{singular value decomposition}\}$
\State $[\bSigma]_{ii}\leftarrow ([\bSigma]_{ii} - \lambda/t)_{+}, i=1, \dots, d$
\State $\bX'\leftarrow \bX \{\text{record previous step}\}$
\State $\bX \leftarrow \mathcal{P}(\bU\bSigma \bV^T)$
\State If $f(\bX) > f(\bX')$, then $t\leftarrow \eta t, \text{go to 4}$.
\State If $|f(\bX) -  f(\bX')| < 10^{-5}$ then exit current iteration and go to the next iteration;
\EndFor
\State $\widehat \bP_{\lambda} =  \bX/\text{sum}(\bX)$.
\EndFor
\end{algorithmic}
\end{algorithm}
In this algorithm, $\mathcal{P}(\bZ) = \frac{I}{\text{sum}(\bZ_{+})}\bZ_{+}$, where $I = n$, the number of samples. Cost function $f(\bX) = \text{is-non-zero}(\bM)\circ \bX - \bM \circ \log(\bX)$.

\subsection{A Proximal Gradient Descent for the Weighted Objective}
\label{sec:solver}
After obtaining a weight matrix which makes the upper error bound small, we plug it into Program ($\ref{weighted-trace-norm}$). Although Program ($\ref{weighted-trace-norm}$) is a convex program and can be solved by casting it into a semidefinite programming (SDP) problem, SDP solvers cannot usually solve the problem when $d_r$ and $d_c$ are both much larger than 100, as discussed in \cite{FPC}. Many iterative methods have been proposed to remedy the large-scale nature of these problems. For example, the algorithm of \cite{FPC} is a proximal gradient method. We adapt their algorithm to our weighted objective by change of variable and we expect their proof technique to be easily applicable to this generalization. In this section, we present our iterative solver which is based on the projected gradient descent method. We first introduce a useful transformation that helps apply the projected gradient descent method.

Consider the following mapping: $\bB \rightarrow \bN:=\sqrt{\bW}\circ \bB$.
Consequently, let the modified observation operator be: $[\tildeopr(\bN)]_i = \langle \bN, \tilde\bX_i \rangle \text{ for all } i \in [n]$, where $\tilde \bX_i = \sqrt{\bW}^{-1} \circ \bX_i$. Note that $\tildeopr(\bN) = \opr(\bB)$ by construction. Then Problem (\ref{weighted-trace-norm}) can be formulated as the following:
\begin{align}
   \min_{\bN\in \mathbb{R}^{d_r\times d_c}} \mathcal{\tilde L}(\bN):=  \frac{1}{n}\|Y - \mathfrak{\tilde X}(\bN)\|_2^2 + \lambda \|\bN\|_*.
 \label{weighted-trace-norm-2}
\end{align}
Notice that Problem (\ref{weighted-trace-norm-2}) resembles the trace-norm penalized regression in Problem (\ref{regular-trace-norm}), which can be solved by the projected gradient descent method to deliver estimator $\widehat\bN$. We can then report the estimator for $\bB^*$ as $\widehat\bB = \sqrt{\bW}^{-1}\circ \widehat\bN$. To proceed with the projected gradient descent method to solve (\ref{weighted-trace-norm-2}), let us denote $g(\bN) =\frac{1}{n}\|Y - \mathfrak{\tilde X}(\bN)\|_2^2$ and $h(\bN) =\lambda \|\bN\|_*$, then $\mathcal{\tilde L}(\bN) = g(\bN) + h(\bN)$. Note that $g$ is a differentiable function whose gradient is $\nabla g(\bN) = \frac{2}{n}\mathfrak{\tilde X}^* (\mathfrak{\tilde X}(\bN) - Y)$, where $\mathfrak{\tilde X}^*$ denotes the dual operator. Recall that the proximal mapping is defined to be:
\[
\text{prox}_{h,t}(\bN) = \argmin_{\bM}\frac{1}{2t}\|\bN - \bM\|_2^2 + h(\bM),
\]
and the generalized gradient of $\mathcal{\tilde L}(\bN)$, denoted as $G_t(\bN)$, is:
\[
G_t(\bN) = \frac{\bN - \text{prox}_{h,t}(\bN - t\nabla g(\bN))}{t}.
\]
In our case, by Lemma \ref{thm:1} proved in Appendix \ref{append:aux}, we have 
\begin{align*}
  \text{prox}_{h,t}(\bN) &= \argmin_{\bM}\frac{1}{2t}\|\bN - \bM\|_2^2 + h(\bM) =S_{\lambda t}(\bN),
\end{align*}
where $S_{\lambda}(\bN)$ is the matrix soft-thresholding operator: $S_{\lambda}(\bN) = \bU \diag{S_{\lambda}(\diag{\bSigma}}\bV^{\top}$ and where $\bN = \bU \bSigma \bV^T$ is the SVD. We write out the proximal-gradient-descent-based Algorithm \ref{alg:fpcw} to solve Problem (\ref{weighted-trace-norm-2}).
\begin{algorithm}[h]
\caption{Iterative solver for weighted matrix completion}
\label{alg:fpcw}
\begin{algorithmic}[1]
\State Initialize $\bN = 0$. Choose step size shrinkage parameter $0<\beta<1$, initial step size $t_{init}$, and tolerance parameter $tol$.
\State Do for tuning parameter $\lambda = \lambda_1 > \lambda_2 >\dots > \lambda_L$:
\begin{enumerate}[label=\alph*)]
    \item Repeat:
    \begin{enumerate}[label=(\roman*)]
    \item Store previous value: $\bN_{old} := \bN$.
        \item Select step size through line search. Initialize $t:= t_{init}$.
        If $$g(\bN - t_{init}G_{t_{init}}(\bN)) > g(\bN) - t_{init} \langle\nabla g(\bN), G_{t_{init}}(\bN)\rangle + \frac{t_{init}}{2} \|G_{t_{init}}(\bN)\|_F^2,$$
        
        then, while $$g(\bN - tG_{t}(\bN)) > g(\bN) - t \langle\nabla g(\bN), G_t(\bN)\rangle + \frac{t}{2} \|G_t(\bN)\|_F^2,$$
        shrink $t:=\beta t$. Else, while 
        $$g(\bN - tG_{t}(\bN)) \leq g(\bN) - t \langle\nabla g(\bN), G_t(\bN)\rangle + \frac{t}{2} \|G_t(\bN)\|_F^2,$$
        enlarge $t:=\frac{1}{\beta} t$. After exiting this forthtracking while loop, return to the second to the last step size $t:=\beta t$.  
        
        \item Update:
        $\bN := \bN - tG_t(\bN)$.
\end{enumerate}
until stopping criterion is satisfied, i.e. $\|\bN_{old} - \bN\|_F^2\leq tol$. 
\item Record $\bN_{\lambda} = \bN$.
\end{enumerate}
\State Output the sequence of solutions $ \sqrt{\bW}^{-1} \circ \bN_{\lambda_1}, \dots, \sqrt{\bW}^{-1}\circ \bN_{\lambda_L}$.
\end{algorithmic}
\end{algorithm}

We have shown how to estimate $\bB^*$ and $\bP^*$ in order to construct the desired weight matrix $\bW$, which is then used in our weighted-trace-norm penalized estimator. We have also outlined a fast iterative method to obtain the estimator. Next, we sketch some proof ideas of the theorem that prompts us to use Program (\ref{weight-construction}) to construct the weight matrix.

\section{Key Steps of the Analysis of Weighted Matrix Completion}
\label{theory}
In this section, we outline the proof strategy for Theorem \ref{matcompthm}. The proof primarily consists of three main steps, and the details of each step are provided in the appendices. Here is a high-level overview of them. First, we begin by proving a deterministic upper bound for the estimation error under some conditions on the data generating processes. Second, we show that these conditions in fact hold with high probability. Third, we combine the two previous results, and with some algebra, derive concrete error bounds with appropriate constant terms.

Before we delve into the deterministic result, we need the following assumption that controls the spikiness of the matrix by bounding $\mathfrak{N}(\sqrt{\bW}\circ\bB^*)$:
\begin{assumption}
\label{assump_spiky}
Assume that $\mathfrak{N}(\sqrt{\bW}\circ\bB^*)\leq n^*$ for some $n^* > 0$.
\end{assumption} 

\paragraph{Step 1 (A deterministic bound).} With the mapping we introduce in Section \ref{sec:solver}, we can directly apply Theorem 3.1 from \cite{nima} to derive the following deterministic result.

 \begin{proposition}[Theorem 3.1 from \cite{nima}]
\label{prop:deterministic}
 Define $\tilde \eta = 72\tr$. Assume that, with constants $\tilde \alpha = \tilde\alpha(\tildeopr)$ and $\tilde \beta = \tilde\beta(\tildeopr)$, for all $\bdelta\in \mathcal{C}(\tilde \nu, \tilde \eta):=\{\bdelta\in \mathbb{R}^{d\times d} \mid \mathfrak{N}(\bdelta) =1, \|\bdelta\|_F \geq \tilde \nu, \|\bdelta\|_* \leq \sqrt{\tilde \eta} \|\bdelta\|_F\}$, we have 
\begin{align}
 \frac{\|\tildeopr(\bdelta)\|_2^2}{n} \geq \tilde \alpha(\tildeopr)\|\bdelta\|^2_F - \tilde \beta(\tildeopr). 
 \label{RSCinprop}
\end{align}

Additionally, we assume that $\tilde\lambda$ is chosen such that 
\begin{align}
    \tilde\lambda\geq 3\|\tilde \bSigma\|_{\op},
    \label{lambdaass}
\end{align} 
where $\tilde \bSigma:=\frac{1}{n}\sum_{i = 1}^n \epsilon_i \tilde \bX_i$. Then for any matrix $\widehat \bB$ such that $\widehat \bN = \sqrt{\bW}\circ \widehat \bB$ satisfies $\mathcal{\tilde L}(\widehat\bN) \leq \mathcal{\tilde L}(\bN^*)$, we have 
\[
\|\widehat \bN - \bN^*\|_F^2 \leq (\frac{100 \tilde\lambda^2 \tr}{3\tilde\alpha^2} + \frac{8{n^*}^2\tilde\beta}{\tilde\alpha}) \vee 4{n^*}^2\tilde\nu.
\]
\end{proposition}

In the above deterministic result, Condition $(\ref{RSCinprop})$ and Condition $(\ref{lambdaass})$ are assumptions that will be proved to hold with high probability with one additional distribution assumption on $\Pi$:

\begin{assumption}
\label{assump_on_frak_c}
There exists $\mathfrak{c}>0$ such that 
\[
\mathbb{E}\Big[\langle \bX, \bB\rangle^2 \cdot \mathbb{I}\big(\|\langle \bX, \bB\rangle\| \leq \mathfrak{c}\big)\Big]\geq \frac{1}{2}\mathbb{E}[\langle \bX, \bB\rangle^2]
\]
for all $\bB$ such that $\mathfrak{N}(\sqrt{\bW}\circ\bB) \leq 1$, where the expectations are with respect to $\Pi$.
\end{assumption}

\paragraph{Step 2 (A probabilistic bound).} Note that Condition $(\ref{RSCinprop})$ is called the restricted strong convexity condition (RSC) satisfied by the observation operator $\tildeopr$ over the set $\mathcal{C}(\tilde \nu, \tilde \eta)$. This condition goes back to the work of \cite{sahand}, and it makes the derivation of the non-asymptotic error bounds for matrix estimation problems possible. Intuitively, if we consider the square loss objective $\frac{1}{2n}\|Y - \tildeopr(\bN)\|_2^2$, then the Hessian matrix of this function is given by $\tildeopr^*\tildeopr/n$, where $\tildeopr^*$ is the adjoint operator of $\tildeopr$. The RSC condition implies that the quadratic loss is strongly convex in a restricted set $\mathcal{C}$ of directions $\bdelta$. That is to say, the observation operator captures a substantial component of a set of matrices whose spikiness and low-rankness are controlled as in $\mathcal{C}(\tilde \nu, \tilde \eta)$. The next result shows that the RSC condition holds with high probability for the modified observation operator $\tildeopr$:

\begin{lemma}[RSC]
\label{lemma:RSC}
With probability greater than $1 - 2\exp(-\frac{Cn\tilde\nu d^2}{\mathfrak{c}^2l})$, the inequality
\begin{align*}
  \frac{\|\tildeopr(\bdelta)\|_2^2}{n} \geq \frac{d^2}{4l}\|\bdelta\|_{F}^2 - 93\frac{l}{d^2}\tilde\eta \mathfrak{c}^2\mathbb{E}[\|\tilde\bSigma_R\|_{\op}]^2
\end{align*}
holds for all $\bdelta \in\mathcal{C}(\tilde\nu, \tilde \eta)$, where $C>0$ is an absolute constant, provided that $Cn\tilde\nu > l\mathfrak{c}^2$, and $\tilde\bSigma_R := \frac{1}{n}\sum_{i=1}^n\zeta_i\tilde\bX_i$ where $\{\zeta_i\}_{i=1}^n$ is an i.i.d. sequence with Rademacher distribution.
\end{lemma}

Now, by denoting the threshold for $\tlambda$ by $\tlambda_1:=C\mathbb{E}\Big[\|\tilde\bSigma_R\|_{\op}\Big]n^*\mathfrak{c}$ and defining
\[
\tilde\alpha:=\frac{d^2}{4l},\hspace{4mm} \tilde\beta:=6696\tilde r\mathfrak{c}^2\frac{l}{d^2}\mathbb{E}\Big[\|\tilde\bSigma_R\|_{\op}\Big]^2, \hspace{4mm}\text{ and }\hspace{4mm} \tilde\nu:= \frac{\tlambda_1^2\tilde rl^2}{{n^*}^2 d^4},
\]
and putting together Proposition \ref{prop:deterministic} and Lemma \ref{lemma:RSC}, we obtain
\begin{align}
   \|\widehat \bN - \bN^*\|_F^2 \leq \Big(\frac{1600\tlambda^2\tilde rl^2}{3d^4} + \frac{32 \times 6696\tlambda_1^2\tilde rl^2}{C^2 d^4}\Big)\vee \frac{4\tlambda^2_1\tilde rl^2}{d^4}\nonumber \leq \frac{C'\tlambda^2\tilde rl^2}{d^4},
\end{align}
for sufficiently large constant $C'>0$. We note that the only condition of Lemma \ref{lemma:RSC} can be shown to hold by taking $C''$ such that $C''n\tlambda^2\tilde r l> \mathfrak{c}^2{n^*}^2 d^2$. Now, we can use Assumption \ref{def_l} to derive the following bound:
\begin{align}
 \|\widehat \bN - \bN^*\|_F^2  = \|\frac{\sqrt{\bW}}{\sqrt{\bP}} \circ \sqrt{\bP}\circ (\widehat\bB - \bB^*)\|_F^2 \geq \frac{1}{ld^2}\|\widehat\bB - \bB^*\|_{L^2(\Pi)}^2.\nonumber
\end{align}
Therefore, when $\tlambda\geq C\mathbb{E}\Big[\|\tilde\bSigma_R\|_{\op}\Big]n^*\mathfrak{c}$,
\begin{align}
  \|\widehat\bB - \bB^*\|_{L^2(\Pi)}^2\leq \frac{C'\tlambda^2\tilde rl^3}{d^2}  
  \label{beforeBernstein}
\end{align}
holds with probability at least $1 - \mathbb{P}(\tlambda<3\|\tilde\bSigma\|_{\op}) - 2\exp(-\frac{C'' n\tilde\lambda^2\tilde r l}{\mathfrak{c}^2{n^*}^2d^2})$.

To obtain a bound for $\mathbb{P}(\tlambda<3\|\tilde\bSigma\|_{\op})$, we rely on a variant of Bernstein tail inequality for the operator norm of matrix martingales, adapted from Lemma 5 in \cite{klopp}, also stated in Appendix \ref{append:bern} for completeness. To specify an explicit choice of $\tilde\lambda$ by bounding the operator norm of $\tilde\bSigma_R$, we use Lemma 6 from \cite{klopp}, also stated in Appendix \ref{append:bern} for completeness.

\paragraph{Step 3 (A concrete bound).} Lastly, we use Orlicz norm to guide selection of $\mathfrak{N}(\bN) = d^2\|\bN\|_{\infty}$ for any $\bB$ and $\bN =\sqrt{\bW}\circ\bB$. So $n^* = 2d^2\|\sqrt{\bW}\circ\bB^*\|_{\infty}$. Consequently, $\mathfrak{c} = \frac{9}{d}\sqrt{\frac{l}{p_{\min}}}$ fulfills Assumption \ref{assump_on_frak_c}. The details are shown in Appendix \ref{appendix:thm1proof}. Building on these quantities and integrating Lemma \ref{corollaryextended1} and \ref{corollaryextended2} into the probabilistic bound $(\ref{beforeBernstein})$, we obtain Theorem \ref{matcompthm}.  

\section{Empirical Results}
\label{sec:emp}
The goal of this section is to compare the performance of our NU-Recommend method with other methods. We perform two sets of experiments, one on synthetic data and the other on real data from two different domains: movie recommendations and lab test recommendations.

\paragraph{Benchmarks.} We compare our NU-Recommend method against (i) regular trace-norm penalized regression that goes back to \cite{plan}, abbreviated as the Uniform method, (ii) marginal-weighted trace-norm penalized regression that goes back to \cite{NIPS2010_4102}, abbreviated as the Margin method, (iii) inverse propensity weighting strategy proposed by \cite{mnar} and applied on the regular trace-norm penalized regression, abbreviated as the IPW+Uniform method, and (iv) modified USVT method designed by \cite{bhattacharya2022matrix}, abbreviated as the ModUSVT method. We estimate marginal probabilities by their empirical estimates. For the last benchmark, although the authors recommended a fixed hyperparameter value, we still tuned the number of singular values for the step of singular value thresholding to further enhance ModUSVT's performance.

\subsection{Synthetic Data}
With the synthetic dataset, our goal is to see how the NU-Recommend algorithm compares against existing benchmarks when we generate ground truth matrix ourselves, which means it could satisfy assumptions such as the low-rankness of the preference matrix and the sampling matrix.

\paragraph{Synthetic Data Generation.}

We assume $d_r = d_c = d = 100$, $r_{\bB} = r_{\bP} = 20$. This means the ground truth preference matrix $\bB^*$ as well as the ground truth sampling matrix $\bP^*$ is 100 by 100 and has rank 20. We use the underlying model $\bB^* = \bU_{\bB}\bV_{\bB}^{\top}$, where $\bU_{\bB}$ and $\bV_{\bB}$ are random matrices of size 100 by 20 with entries drawn independently and uniformly from $[0,1]$. Similarly, we let the sampling matrix $\bP^*$ to be $\bU_{\bP}\bV_{\bP}^{\top}$, where $\bU_{\bP}$ and $\bV_{\bP}$ are random matrices of size 100 by 20 with entries drawn independently and uniformly from $[0,1]$. We then normalize the entries of the sampling matrix to satisfy the sampling model. We set the noise variance to be $\sigma^2 = 1$. We try different sample sizes $n$ from $\{1000, 1200, 1400, 1600, 1800, 2000\}.$ We generate 100 datasets for each sample size and compare the relative estimation errors (with 2 SE error bars) in both Frobenius norm and $L_2(\Pi)$ norm for all algorithms across these 100 runs. We look at a comprehensive range of $\lambda$'s for all algorithms and for each algorithm we pick the lambda that provides the lowest estimation error. We do not perform cross-validation here because we already know the ground truth matrix. Later, when we deal with the real data, we will be using the cross-validation procedure to select a $\lambda$. As we have mentioned in Section \ref{weight-mat-constr}, NU-Recommend needs a raw $\widehat\bB$ from another estimator as an input and also an estimation on the sampling matrix $\widehat\bP$. We take the raw $\widehat\bB$ from the Margin method as it empirically performs better than the Uniform method. In Section \ref{sec:samp-est}, we point out that $\widehat\bP$ could be obtained from a Poisson matrix recovery. Alternatively, one can take a simpler route and use the marginal probabilities. We pick the latter because it is computationally faster.

\begin{figure}
    \centering
    \includegraphics[scale=0.6]{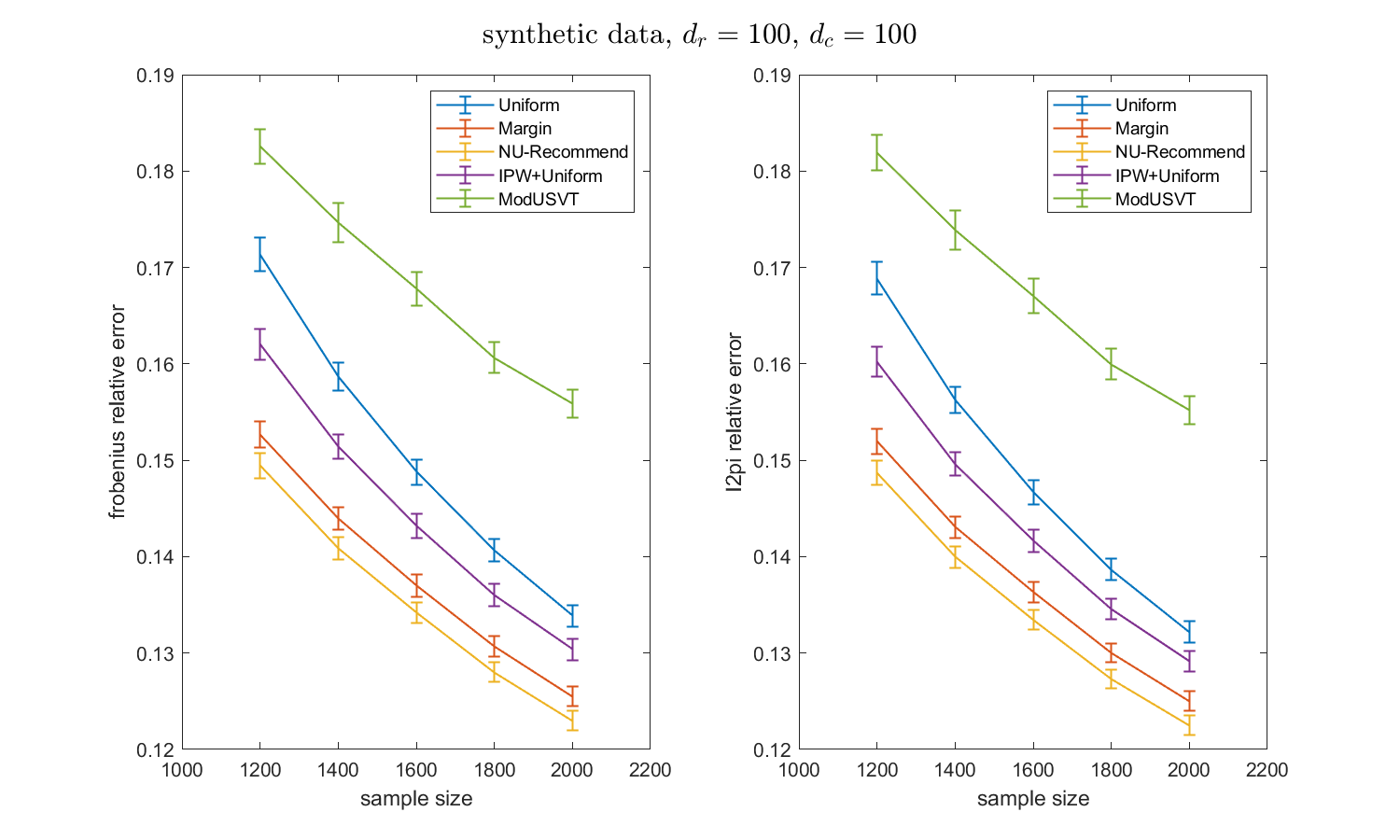}
    \caption{plots of the relative frobenius and $L_2(\Pi)$ errors versus sample sizes 1000, 1200, 1400, 1600, 1800, 2000}
    \label{fig:plot_synthetic}
\end{figure}

\paragraph{Results of synthetic experiment.} Figure \ref{fig:plot_synthetic} shows that our method NU-Recommend gives the lowest relative matrix estimation error both in Frobenius norm and in $L_2(\Pi)$ norm among all methods. On average, 10.18\% better than the IPW, 6.53\% better than the Uniform and 2.08\% better than the Margin. ModUSVT performs much worse than all other methods. This is because their method focuses on improving computation time and not the statistical efficiency. \cite{agarwal2021causal} also reported that ModUSVT does not perform well in their setting. Since ModUSVT cannot handle repeated observations for an entry, we take the average observed values and for each entry and treat the average as its observation.

\subsection{MovieLens Dataset}
\begin{figure}
\centering
\includegraphics[scale = 0.8]{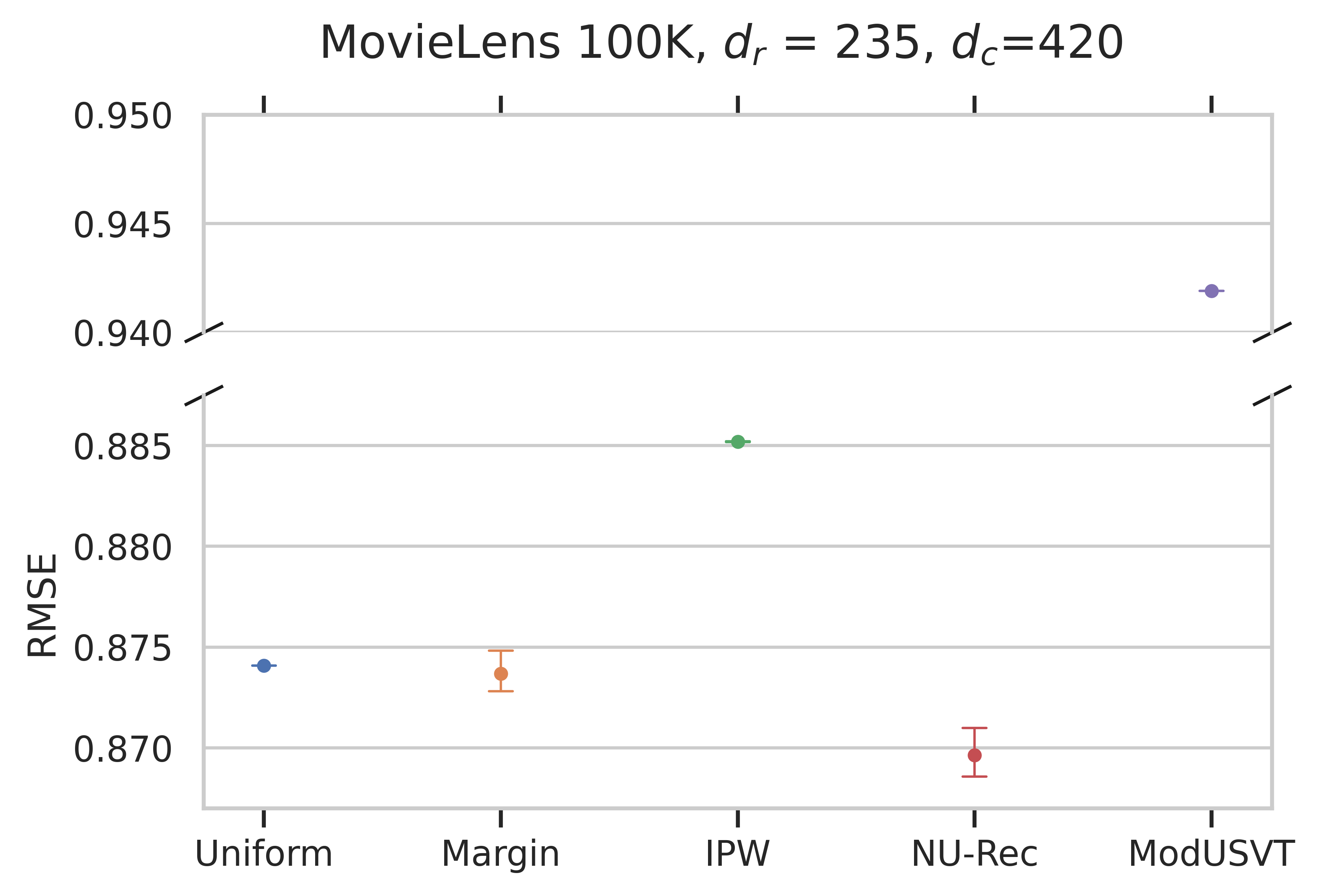}
\caption{The matrix dimension is $235\times 420$, NU-Recommend is taken at $L = 3, \gamma = 3$, 20 runs. The range of 0.885 to 0.940 is removed for better display.}
\label{fig:MovieLens100k_25_wUSVT}
\end{figure}

\paragraph{Sub-sample of MovieLens 100K dataset.} In addition to synthetic experiments, we also perform experiments using real-world data to show the practicality of our method. Specifically, the low-rank assumption on both $\bB^*$ and $\bP^*$ may not necessarily hold. The first dataset we use, MovieLens 100K, is a common collaborative filtering benchmark dataset. This dataset contains 100,000 ratings from 943 users on 1682 movies. Each user has rated at least 20 movies. We work with a dense sub-matrix of 235 users, 420 movies, and 40,000 observations for the illustration purpose. This matrix is constructed by taking the observations only from the top 25\% users who rated the most and the top 25\% movies that were rated the most. With this matrix, we first randomly select 80\% samples as evaluation and 20\% samples as testing. We further partition the evaluation set to an 80/20 training and validation split and use that to optimize $\lambda$. Once the optimal $\lambda$ is selected, we refit the algorithm using all of the evaluation set, and show results on the test set. We repeat this process by 20 splits of the evaluation set. This leads to 20 RMSE values for each algorithm on the test set. We report these RMSE with 2 standard errors in Figure \ref{fig:MovieLens100k_25_wUSVT}.

\paragraph{Results of MovieLens.} Figure \ref{fig:MovieLens100k_25_wUSVT} shows that our method NU-Recommend gives the lowest test RMSE among all methods. As we can see, NU-Recommend improves RMSE by 1.76\% compared with the IPW+Uniform method, by 0.51\% compared with the Uniform method, and by 0.45\% compared with the Margin method. While these improvements may seem small, they are substantial for this problem. For example, during the final year of the Netflix Prize competition, the best team's RMSE moved from 0.8627 to 0.8567 (less than 0.7\% improvement).

\subsection{Lab Test Dataset}
\paragraph{Sub-sample of lab test dataset.} The other real dataset we use contains data for patient visits at Stanford Hospital and Clinics. In this dataset, we have access to 54653 patients in total and up to 4559 lab tests taken over one year period. If a patient has multiple lab tests in that one year, we will just use the average value. Therefore, we have a matrix of 54653 number of rows and 4559 number of columns. If a patient does not have a test, then the corresponding entry has a missing value. This matrix has many missing entries, hence we restrict the data to those patients who have more than 110 lab test results, and to those lab tests which are done by at least 1000 patients. 

This leads to a matrix completion problem with 259 patients (rows), 198 tests (columns) and about 40\% of the matrix entries are observed. The lab test results are all positive real numbers, and the distribution is skewed, so we take a $\log(1+x)$ transformation on the entries and standardize the columns. We cross-validate for $\lambda$ the same way as we have done for the MovieLens data. We repeat the process by 10 splits of the evaluation set.

\begin{figure}
    \centering
    \includegraphics[scale = 0.8]{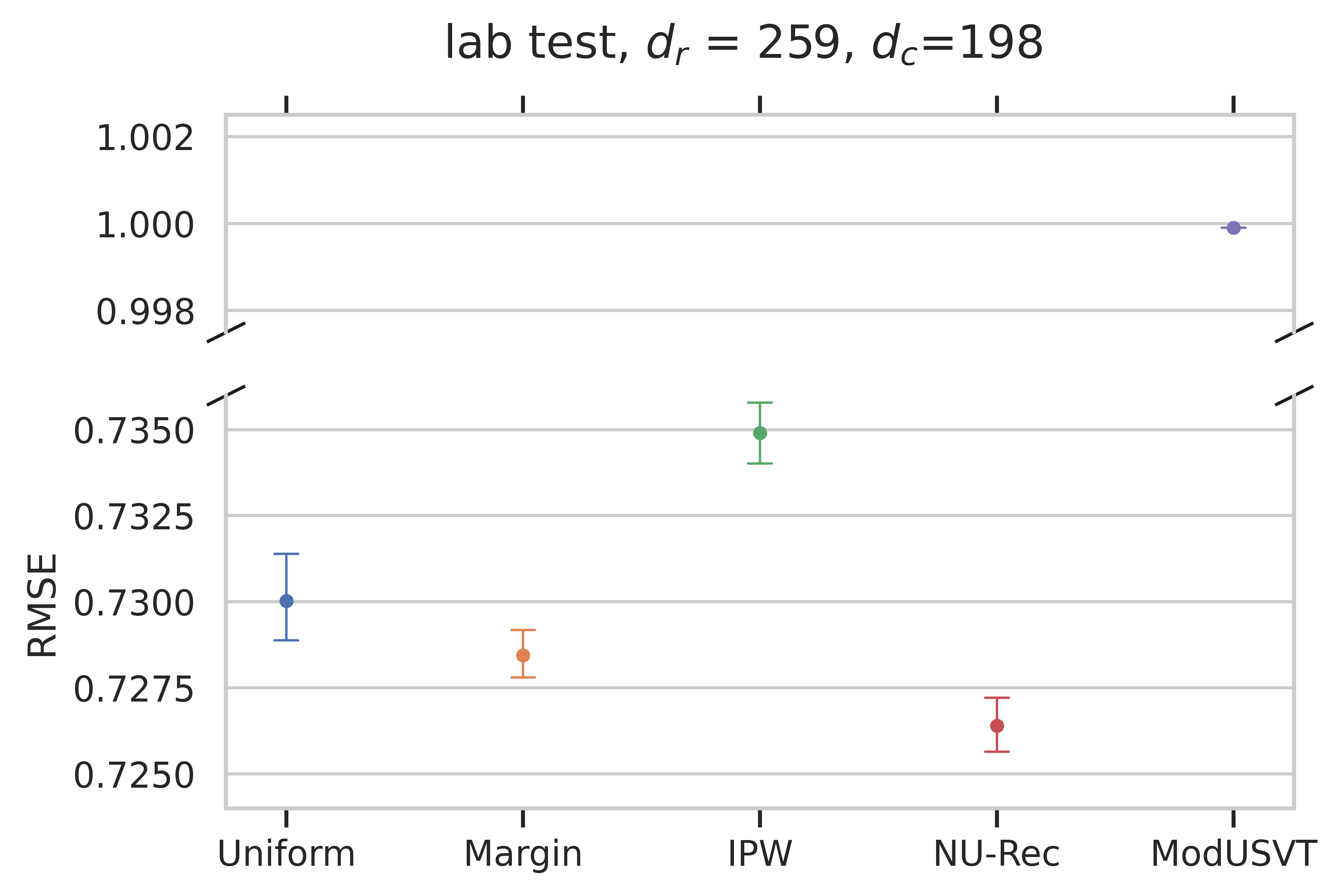}
    \caption{The matrix dimension is $259\times 198$, NU-Recommend is taken at $L = 3, \gamma = 3$, 10 runs. The range of 0.7350 to 0.998 is removed for better display.}
    \label{fig:lab-test}
\end{figure}

\paragraph{Results of lab test data.} We present the results in Figure \ref{fig:lab-test} in the same way as we did for MovieLens. As we can see, our method NU-Recommend gives the lowest test RMSE among all methods. NU-Recommend improves RMSE by 1.16\% compared with the IPW+Uniform method, by 0.50\%, compared with the Uniform method, and by 0.28\%, compared with the Margin method. Overall, these results are consistent with our result on MovieLens Data. We have also run ModUSVT on the lab test data and it cannot emulate the performance of NU-Recommend. ModUSVT needs interval of the entries as input, so we use the maximum and minimum values of the training lab test data to set the range.

\subsection{Restoring Fairness with Non-uniformity}
\paragraph{Synthetic data reveals fair predictions of NU-Recommend.} We first work with synthetic data where the data missing pattern is random (MCAR to be precise), and show that our algorithm NU-Recommend not only delivers better overall RMSE, but also predicts better on the rows with fewer observations, restoring fairness by not over-obsessing with accurately predicting entries of rows with more observations. Specifically, we examine whether the prediction algorithms bring similar prediction qualities for all rows, regardless of their chance of being observed. Even though our goal is to bring the more realistic non-uniformity of the data into our method to lower the overall RMSE, our algorithm allows for more equitable distribution of prediction error. On a high level, by using a weighted penalization scheme, we give different weights to rows providing different amount of information, thus balancing prediction errors across rows of different sampling probabilities, despite the actual more involved mechanism derived from the upper bounds.

We regress per-row prediction error on the estimated probability of observing each row. Intuitively, the prediction for rows that have more observations should be better than the prediction for rows that have fewer observations, since we have more data for the former rows. With the uniform method, we indeed notice that the accuracy for rows that have more observations is better than those with fewer observations (the coefficient for the regressor is negative and the $p$-value for the coefficient is less than 0.05), showing that the model indeed focuses more on rows with more observations. However, there is no statistically significant relationship between NU-Recommend accuracy and the estimated probability of rating ($p$-value is more than 0.05). The same analysis holds when we regress per-column prediction error on the estimated probability of observing each column.

\paragraph{MovieLens data validates the fairness analysis.} Next, we show that our NU-Recommend algorithm brings similar prediction qualities for all users for the MovieLens dataset, regardless of their frequency of rating, and thus restores fairness by not over-obsessing with accurately predicting preferences of highly active users. We want to point out that our regression analysis assumes MCAR which is not as realistic as modeling MovieLens data with MNAR, as the data are observational. Additional analysis that incorporates covariates to control for the non-random missing pattern is left as future work.

We follow the same procedures that we perform for the synthetic data, and regress per-user prediction error on the estimated probability of observing each user's rating. Intuitively, the prediction for users who rate a lot should be better than the prediction for users who rate much less, since we have more data for the former users. With the uniform method, we indeed notice that the accuracy for users who rate more is better than those who rate less (the coefficient for the regressor is negative and the $p$-value for the coefficient is less than 0.05), showing that the model indeed focuses more on active users. However, there is no statistically significant relationship between NU-Recommend accuracy and the estimated probability of rating ($p$-value is more than 0.05). 

More accurate predictions lead to more accurate product recommendations and better quality recommendations. With the uniform method, quality of recommendations is significantly better for users with more ratings, but NU-Recommend balances quality between users with different rating frequencies. We want to point out that providing high quality recommendations across different users is not the only notion of fairness. But if this is the goal, then NU-Recommend allows it. It would be an important future direction to design algorithms that tailor to different fairness criteria.

\paragraph{Bias and variance tradeoff.} The mathematical component underpinning the fairness mechanism is the bias and variance tradeoff. The fairness lies at equating the sum of variance and bias across all entries. Our NU-Recommend method makes sure that biasing via the regularization term is done more fairly than the Uniform method. More active users have a lower variance since there are more terms in the quadratic loss for users with more ratings. Vice versa, infrequent users have a higher variance due to fewer terms in the quadratic loss, which leads to overfitting. By assigning less weight to the entries with low sampling rate inside the nuclear norm penalization, NU-Recommend biases active users more than the infrequent users, since active users afford to have high bias given their low variance. Uniform method does not account for the different levels of variance and does not correct for the unfair biases. In fact, IPW+Uniform method proposed in \cite{mnar} also makes use of this bias-variance tradeoff mechanism by balancing the variances, while NU-Recommend method balances biases. Furthermore, for example in healthcare, works such as \cite{seyyed2021underdiagnosis} have shown that state-of-the-art AI-based algorithms would amplify human biases and under-diagnose under-represented groups. We believe that carefully considering the bias and variance tradeoff in machine learning based healthcare systems is essential to treat historically under-served patient populations equally and fairly.

\section{Conclusions}
This paper introduced a new method to learn user preferences for recommender systems. Motivated by the non-uniform pattern of observing user preferences, we formulated a general weighted-trace-norm penalized regression and examined the upper bound on its estimation error. Experiments on both synthetic data and real data MovieLens and lab test provided empirical evidence that recovering the non-uniformity is beneficial. Furthermore, we showed that, by selecting the weight matrix that minimizes the upper bound, the performance can be further enhanced. 

A number of exciting future research are as follows. Our theoretical upper bound is motivated by the oracle inequalities in high-dimensional statistics. It is interesting to see that we could take this upper bound and then from it derive a better version of the weighted estimator, which empirically performs very well. We believe this direction could be further explored for the high-dimensional statistics literature. 

One practical extension is to explore how to select the weight matrix more efficiently. For some applications such as healthcare industry, the training process needs to be done faster. With the current convex program solved by semidefinite programming to obtain the desired weight matrix, the scalability of our approach can be compromised. Nonetheless, the focus of this paper is to show that statistical efficiency is achieved by incorporating the non-uniform observation pattern in a theory-driven way. How to improve computation efficiency is left for future work. 

Side information in the form of observed user and item context can be incorporated to help learn the preference matrix. It would be interesting to combine our method with works such as \cite{chiang2015matrix,yu2022nonparametric}, which built on traditional matrix completion methods and adds contexts, so that observed contexts can be leveraged in the estimation procedure.

Another important direction is to include side information to aid the sampling matrix estimation. For example, in the MovieLens case, the rating data can be much more sparse than the watch data. Thus, to recover the sampling scheme of the rating data, we can instead change to recover that of the watch data. Although the sampling matrix for rating data and watch data may not be exactly the same, we can model the degree to which they reflect each other to make a better informed estimation. 

It would also be very valuable to be able to conduct inference on the estimated preference matrix. Works such as \cite{chen2019inference} provided recipes for existing matrix completion methods on how to quantify uncertainty and identify a short interval that is likely to contain a missing entry. To combine existing works with our method is a promising future direction.

Last but not least, as mentioned before, it is worth exploring how to extend the work to the ``observational data regime'' where the missing pattern of an entry may depend on the true value of that entry. For example, in the healthcare setting, patients get tested based on their symptoms, so the sampling probability may depend on the testing results. However, as we have shown earlier, our algorithm can perform better than an algorithm that models data-dependent missing pattern. Therefore, how to model the dependency more efficiently requires careful thinking.

\bibliography{refs}
\bibliographystyle{ormsv080}

\begin{APPENDICES}
\section{Proof of Proposition \ref{prop:deterministic}}

First, it follows from $\mathcal{\tilde L}(\bN^*) > \mathcal{\tilde L}(\widehat\bN) $ that
\[
\frac{1}{n}\|Y - \tildeopr(\widehat \bN)\|_2^2 + \tlambda \|\widehat \bN\|_*\leq \frac{1}{n}\|Y - \tildeopr(\bN^*)\|_2^2 + \tlambda \|\bN^*\|_*. 
\]
By substituting $Y$ with $\tildeopr(\bN^*) + E$ and doing some algebra, we have 
\[
\frac{1}{n}\|\tildeopr(\bN^* - \widehat\bN)\|_2^2 + 2\langle\tilde \bSigma, \bN^* - \widehat \bN\rangle + \tlambda \|\widehat \bN\|_* \leq \tlambda \|\bN^*\|_*.
\]

Then, using duality between the operator norm and the trace norm, we get 
\[
\frac{1}{n}\|\tildeopr(\bN^* - \widehat\bN)\|_2^2+ \tlambda \|\widehat \bN\|_* \leq 2\|\tilde\bSigma\|_{\op}\cdot \|\bN^* - \widehat\bN\|_*+ \tlambda \|\bN^*\|_*.
\]

For a given set of vectors $S$, we denote by $\bP_{S}$ the orthogonal projection on the linear subspace spanned by elements of $S$ (i.e., $\bP_{S}=\sum_{i=1}^k u_iu_i^{\top}$ if $\{u_1, \dots, u_k\}$ is an orthogonal basis for $S$). For matrix $\bN\in \mathbb{R}^{d_r\times d_c},$ let $S_r(\bN)$ and $S_c(\bN)$ be the linear subspace spanned by the left and right orthonormal singular vectors of $\bN$, respectively. Then, for $\bA\in \mathbb{R}^{d_r\times d_c}$ define 
\[
\bP^{\perp}_{\bN}(\bA):=\bP_{S^{\perp}_r(\bN)}\bA \bP_{S^{\perp}_c(\bN)} \text{ and } \bP_{\bN}(\bA):=\bA - \bP^{\perp}_{\bN}(\bA).
\]

We can alternatively express $\bP_{\bN}(\bA)$ as 
\begin{align}
  \bP_{\bN}(\bA) =\bP_{S_r(\bN)}\bA + \bP_{S^{\perp}_r(\bN)}\bA \bP_{S_c(\bN)}.
  \label{alternate}
\end{align}

In particular, since $S_r(\bN)$ and $S_c(\bN)$ both have dimension rank($\bN$), it follows from (\ref{alternate}) that 
\[
\rank(\bP_{\bN}(\bA))\leq 2\rank(\bN).
\]
Moreover, the definition of $\bP_{\bN^{\perp}}$ implies that the left and right singular vectors of $\bP_{\bN^{\perp}}(\bA)$ are orthogonal to those of $\bN$. We thus have
\[
\|\bN+\bP_{\bN}^{\perp}(\bA)\|_* = \|\bN\|_* + \|\bP_{\bN}^{\perp}(\bA)\|_*.
\]

By setting $\bN:=\bN^*$ and $\bA:=\hat\bN - \bN^*$, the above equality entails:
\[
\|\bN^* + \bP^{\perp}_{\bN^*}(\bN^* - \widehat\bN)\|_* = \|\bN^*\|_* + \|\bP^{\perp}_{\bN^*}(\bN^* - \widehat\bN)\|_*.
\]

We can then use the above to get the following inequality:
\begin{align}
   \|\widehat\bN\|_* &=  \|\bN^* + \widehat\bN - \bN^* \|_* \nonumber \\
   &=\|\bN^* + \bP^{\perp}_{\bN^*}(\bN^* - \widehat\bN) + \bP_{\bN^*}(\bN^* - \widehat\bN)\|_* \nonumber\\
   &\geq \|\bN^* + \bP^{\perp}_{\bN^*}(\bN^* - \widehat\bN)\|_* - \|\bP_{\bN^*}(\bN^* - \widehat\bN)\|_* \nonumber\\
   &=\|\bN^*\|_* + \|\bP^{\perp}_{\bN^*}(\bN^* - \widehat\bN)\|_* - \|\bP_{\bN^*}(\bN^* - \widehat\bN)\|_* \label{func3}.
\end{align}
Thus, we get 
\begin{align*}
    \frac{1}{n}\|\tildeopr(\bN^* - \widehat\bN)\|_2^2 &\leq 2\|\tilde\bSigma\|_{\op}\cdot \|\bN^* - \widehat\bN\|_*+ \tlambda\|\bP^{\perp}_{\bN^*}(\bN^* - \widehat\bN)\|_* - \tlambda\|\bP_{\bN^*}(\bN^* - \widehat\bN)\|_* \\
    &\leq (2\|\tilde\bSigma\|_{\op} +\tlambda)\|\bP_{\bN^*}(\bN^* - \widehat\bN)\|_* + (2\|\tilde\bSigma\|_{\op} -\tlambda)\|\bP^{\perp}_{\bN^*}(\bN^* - \widehat\bN)\|_* \\
    &\leq \frac{5}{3}\tlambda \|\bP_{\bN^*}(\bN^* - \widehat\bN)\|_*.
\end{align*}
Now, by the fact that $\rank(\bP_{\bN^*}(\bN^* - \widehat\bN))\leq 2\rank(\bN^*)$, we can apply Cauchy-Schwartz to singular values of $\bP_{\bN^*}(\bN^* - \widehat\bN)$ to obtain:
\begin{align*}
 \frac{1}{n}\|\tildeopr(\bN^* - \widehat\bN)\|_2^2&\leq \frac{5}{3}\tlambda \sqrt{2\rank(\bN^*)}\|\bP_{\bN^*}(\bN^* - \widehat\bN)\|_F   \\
 & \leq \frac{5}{3}\tlambda \sqrt{2\rank(\bN^*)}\|\bN^* - \widehat\bN\|_F.
\end{align*}

Next, we want to make a connection between $\widehat \bN$ and $\mathcal{C}(\tilde\nu, \tilde\eta)$ with the following lemma.

\begin{lemma}
\label{lemma_n_c}
If $\tlambda \geq 3\|\tilde \bSigma\|_{\op}$, then
\[
\|\bP^{\perp}_{\bN^*}(\bN^* - \widehat\bN)\|_*\leq 5\|\bP_{\bN^*}(\bN^* - \widehat\bN)\|_*.
\]
\end{lemma}
\proof{Proof of Lemma \ref{lemma_n_c}.}
Note that $\|\tildeopr(\cdot) - Y\|_2^2$ is a convex function. We can then use the convexity at $\bN^*$ to get:
\begin{align*}
   &\frac{1}{n}\|\tildeopr(\widehat\bN) - Y\|_2^2 - \frac{1}{n}\|\tildeopr(\bN^*) - Y\|_2^2 \\
   &\geq -\frac{2}{n}\sum_{i=1}^n(y_i - \langle \tilde \bX_i, \bN^*\rangle)\langle \tilde \bX_i,\widehat\bN - \bN^* \rangle \\
   &=-2\langle\tilde\bSigma, \widehat\bN - \bN^*\rangle \geq -2\|\tilde\bSigma\|_{\op}\|\widehat\bN - \bN^*\|_*\\
   &\geq -\frac{2}{3}\tlambda\|\widehat\bN - \bN^*\|_*.
\end{align*}

We detail the above procedures below (readers who are familiar with convex functions can omit this):

Let $f(\bN) = \frac{1}{n}\|\tildeopr(\bN) - Y\|_2^2$. Then $$[\nabla f(\bN)]_{jk} = \sum_{i=1}^n\frac{2}{n}\tilde\bX_{i, jk}(\langle \tilde \bX_i, \bN^*\rangle - y_i),\text{ so }\nabla f(\bN) = \sum_{i=1}^n\frac{2}{n} \tilde\bX_i(\langle \tilde\bX_i, \bN^*\rangle - y_i).$$
Now by using convexity at $\bN^*$, we get:
\begin{align*}
    f(\widehat\bN) - f(\bN^*) &\geq \langle \nabla_{\bN^*}f(\bN^*), \widehat\bN - \bN^* \rangle\\
    &= \langle \frac{2}{n}\sum_{i=1}^n \tilde\bX_i(\langle \tilde\bX_i, \bN^*\rangle - y_i), \widehat\bN - \bN^*\rangle \\
    &=\frac{2}{n}\sum_{i=1}^n(\langle\tilde\bX_i, \bN^*\rangle - y_i)\langle\widehat\bN - \bN^*, \tilde\bX_i\rangle\\
    &=-\frac{2}{n}\sum_{i=1}^n(y_i-\langle  \tilde\bX_i, \bN^*\rangle)\langle  \tilde\bX_i, \widehat\bN - \bN^*\rangle.
\end{align*}
Combining this, we get, 
\begin{align*}
    \frac{2}{3}\tlambda\|\widehat\bN - \bN^*\|_* &\geq \frac{1}{n}\|\tildeopr(\bN^*) - Y\|_2^2 - \frac{1}{n}\|\tildeopr(\widehat\bN) - Y\|_2^2 \\
    &\geq \tlambda\|\widehat\bN\|_* - \tlambda\|\bN^*\|_* \\
   & \geq \tlambda \|\bP^{\perp}_{\bN^*}(\bN^* - \widehat\bN)\|_* - \tlambda\|\bP_{\bN^*}(\bN^* - \widehat\bN)\|_*.
\end{align*}
Using the triangle inequality, we have
\[
\|\bP^{\perp}_{\bN^*}(\bN^* - \widehat\bN)\|_*\leq 5\|\bP_{\bN^*}(\bN^* - \widehat\bN)\|_*.
\]
\QED
\endproof

Lemma \ref{lemma_n_c}, the triangle inequality and the fact that $\rank(\bP_{\bN^*}(\widehat\bN -\bN^*)) \leq 2 \rank(\bN^*)$ imply that
\begin{align*}
   \|\widehat\bN -\bN^*\|_* &\leq 6 \|\bP_{\bN^*}(\bN^* - \widehat\bN)\|_*\\
   &\leq \sqrt{72 \rank(\bN^*)}\|\bP_{\bN^*}(\bN^* - \widehat\bN)\|_F\\
   &\leq \sqrt{72 \rank(\bN^*)}\|\bN^* - \widehat\bN\|_F.
\end{align*}
By the above, we have made a connection with $\mathcal{C}(\tilde\nu, \tilde\eta)$.

Next, define $\hat n:=\mathfrak{N}(\widehat \bN - \bN^*)$ and $\bA:=\frac{1}{\hat n}(\widehat \bN - \bN^*)$. We then have that 
\[
\mathfrak{N}(\bA) = 1 \text{ and } \|\bA\|_*\leq \sqrt{72\rank(\bN^*)}\|\bA\|_F.
\]
Now, we consider the following two cases:

Case 1: If $\|\bA\|_F^2<\tilde \nu$, then
\[
\|\bN^* - \widehat\bN\|_F^2<4{n^*}^2\tilde\nu.
\]

Case 2: Otherwise, $\bA\in \mathcal{C}(\tilde\nu, \tilde\eta)$. We can, now, use the RSC condition to get 
\[
\tilde\alpha\frac{\|\bN^* - \widehat\bN\|_F^2}{\hat n^2} - \tilde\beta \leq \frac{\|\tildeopr(\bN^* - \widehat\bN)\|_F^2}{n\hat n^2}
\]
which leads to
\begin{align*}
    \tilde\alpha\|\bN^* - \widehat\bN\|_F^2 - {4n^*}^2\tilde\beta &\leq \frac{\|\tildeopr(\bN^* - \widehat\bN)\|_F^2}{n}\\
    &\leq \frac{5\tlambda\sqrt{2\rank(\bN^*)}}{3}\|\bN^* - \widehat\bN\|_F\\
    &\leq \frac{50\tlambda^2\rank(\bN^*)}{3\tilde\alpha} + \frac{\tilde\alpha}{2}\|\bN^* - \widehat\bN\|_F.
\end{align*}
Therefore, we have 
\[
\|\bN^* - \widehat\bN\|_F^2\leq \frac{100\tlambda^2\rank(\bN^*)}{3\tilde\alpha^2} + \frac{{8n^*}^2\tilde\beta}{\tilde\alpha},
\]
which completes the proof of this proposition.
\QED
\endproof

\section{Proof of Lemma \ref{lemma:RSC}}

Next, we will show that the RSC condition holds with high probability. Instead of proving Lemma \ref{lemma:RSC} directly, we show a variant of it first:

\begin{lemma}[Restricted Strong Convexity]
\label{RSC}
Define 
\[
 \mathcal{C'}(\tilde \theta, \tilde \eta):=\{\bdelta\in \mathbb{R}^{d_r\times d_c} | \mathfrak{N}(\bdelta) =1, \|\sqrt{\bW}^{-1}\circ\bdelta\|_{L^2(\Pi)}^2 \geq \tilde\theta, \|\bdelta\|_* \leq \sqrt{\tilde \eta} \|\bdelta\|_F\}.
\]
If Assumption \ref{assump_on_frak_c} hold, then the inequality
\[
\frac{\|\tildeopr(\bdelta)\|_2^2}{n} \geq \frac{1}{4}\|\sqrt{\bW}^{-1}\circ\bdelta\|_{L^2(\Pi)}^2 - 93\frac{l}{d^2}\tilde\eta \mathfrak{c}^2\mathbb{E}[\|\tilde\bSigma_R\|_{\op}]^2
\text{ for all } \bdelta\in \mathcal{C'}(\tilde \theta, \tilde \eta)
\]
holds with probability greater than $1-2\exp(-\frac{Cn\tilde\theta}{\mathfrak{c}^2})$ where $C>0$ is an absolute constant, provided that $Cn\tilde\theta >\mathfrak{c}^2$, and $\tilde\bSigma_R:=\frac{1}{n}\Sigma_{i=1}^{n}\zeta_i\tilde\bX_i$ where $\{\zeta_i\}_{i=1}^n$ is an i.i.d sequence with Rademacher distribution.

\end{lemma}

\proof{Proof of Lemma \ref{RSC}(RSC).}
Set \[
\beta = 93\frac{l}{d^2}\tilde\eta \mathfrak{c}^2\mathbb{E}[\|\tilde\bSigma_R\|_{\op}]^2.
\]
We denote the bad event $\mathcal{B}$ as the following:
\[
\mathcal{B} = \Big\{\exists \bdelta \in \mathcal{C}'(\tilde \theta, \tilde \eta) \text{ such that } \frac{1}{2} \|\sqrt{\bW}^{-1}\circ\bdelta\|_{L^2(\Pi)}^2 - \frac{1}{n}\|\tildeopr(\bdelta)\|_2^2 > \frac{1}{4}\|\sqrt{\bW}^{-1}\circ\bdelta\|_{L^2(\Pi)}^2 + \beta \Big\}.
\]
We thus need to bound the probability of this event. Set $\xi = 6/5.$ Then, for $T>0$, we define
\[
\mathcal{C}'(\tilde \theta, \tilde \eta, T):= \{\bdelta\in \mathcal{C}'(\tilde \theta, \tilde \eta) \mid T\leq \|\sqrt{\bW}^{-1}\circ\bdelta\|_{L^2(\Pi)}^2<\xi T\}.
\]
Clearly, we have 
\[
\mathcal{C}'(\tilde \theta, \tilde \eta) = \cup_{k=1}^{\infty} \mathcal{C}'(\tilde \theta, \tilde \eta, \xi^{k-1}\tilde \theta).
\]
Now, if the event $\mathcal{B}$ holds for some $\bdelta\in \mathcal{C}'(\tilde \nu, \tilde \eta)$, then $\bdelta\in \mathcal{C}'(\tilde \nu, \tilde \eta,\xi^{k-1}\tilde \nu)$ for some $k\in \mathbb{N}$. In this case, we have 
\begin{align*}
\frac{1}{2} \|\sqrt{\bW}^{-1}\circ\bdelta\|_{L^2(\Pi)}^2 - \frac{1}{n}\|\tildeopr(\bdelta)\|_2^2  &> \frac{1}{4}\|\sqrt{\bW}^{-1}\circ\bdelta\|_{L^2(\Pi)}^2 + \beta \\
    &\geq \frac{1}{4} \xi^{k-1}\tilde \theta + \beta\\
    &=\frac{5}{24}\xi^k\tilde \theta +\beta.
\end{align*}

Next, we define the event $\mathcal{B}_k$ as 
\[
\mathcal{B}_k = \Big\{\exists \bdelta \in \mathcal{C}'(\tilde \nu, \tilde \eta, \xi^{k-1}\tilde \nu) \text{ such that } \frac{1}{2} \|\sqrt{\bW}^{-1}\circ\bdelta\|_{L^2(\Pi)}^2 - \frac{1}{n}\|\tildeopr(\bdelta)\|_2^2 > \frac{5}{24}\xi^{k}\tilde \theta + \beta \Big\}.
\]
It follows that 
\[
\mathcal{B}\subseteq \cup_{k=1}^{\infty}\mathcal{B}_k.
\]
The following lemma helps us control the probability that each of these $\mathcal{B}_k$'s happen.

\begin{lemma} 
\label{lemma1}
Define 
\[
Z_T := \sup_{\bdelta \in \mathcal{C}'(\tilde \theta, \tilde \eta,T)}\Big\{\frac{1}{2} \|\sqrt{\bW}^{-1}\circ\bdelta\|_{L^2(\Pi)}^2 - \frac{1}{n}\|\tildeopr(\bdelta)\|_2^2\Big\}.
\]
Then, assuming that $(\bX_i)_{i=1}^n$ are i.i.d. samples drawn from $\Pi$, we get 
\begin{align}
\label{prob1}
\mathbb{P}(Z_T \geq \frac{5\xi T}{24} + \beta)\leq \exp\Bigg(-\frac{Cn\xi T}{\mathfrak{c}^2}\Bigg),
\end{align}
for some numerical constant $C>0$.
\end{lemma}
\proof{Proof of Lemma \ref{lemma1}.}
For a $d_r\times d_c$ matrix $\bdelta$, define
\[
f(\tilde\bX; \bdelta) := \langle\tilde\bX,\bdelta \rangle^2 \cdot \mathbb{I}(|\langle\tilde\bX, \bdelta\rangle|\leq \mathfrak{c}).
\]
Next, letting 
\[
W_T:= \sup_{\bdelta \in \mathcal{C}'(\tilde \theta, \tilde \eta,T)} \frac{1}{n}\sum_{i=1}^n\{\mathbb{E}[f(\tilde \bX_i;\bdelta)] - f(\tilde \bX_i;\bdelta)\},
\]

\[
\widetilde W_T:= \sup_{\bdelta \in \mathcal{C}'(\tilde \theta, \tilde \eta,T)} \Big|\frac{1}{n}\sum_{i=1}^n\{\mathbb{E}[f(\tilde \bX_i;\bdelta)] - f(\tilde \bX_i;\bdelta)\}\Big|,
\]
it follows from Assumption \ref{assump_on_frak_c} (where $\mathfrak{c}$ is defined) that $Z_T \leq W_T$, and clearly $W_T\leq \widetilde W_T$ hence 
\[
\mathbb{P}(Z_T\geq t)\leq \mathbb{P}(\widetilde W_T\geq t),
\]
for all $t$. Therefore, if we prove (\ref{prob1}) holds when $Z_T$ is replaced with $\widetilde W_T$, we would be done. In the remaining, we will aim to prove this via Massart's inequality. In order to invoke Massart's inequality, we need bounds for $\mathbb{E}[\widetilde W_T]$ and $\Var(\widetilde W_T)$.

First, we find an upper bound for $\mathbb{E}[\widetilde W_T]$. It follows from the symmetrization argument that

\begin{align}
\label{ineq1}
  \mathbb{E}[\widetilde W_T] \leq 2\mathbb{E}\Bigg[\sup_{\bdelta \in \mathcal{C}'(\tilde \theta, \tilde \eta,T)} \Big|\frac{1}{n}\sum_{i=1}^n \zeta_i f(\tilde \bX_i;\bdelta)\Big|\Bigg].  
\end{align}

Next, we will use the contraction inequality. First, we write $f(\tilde\bX_i; \bdelta) = \alpha_i \langle\tilde\bX_i; \bdelta\rangle$, where $\alpha_i =\langle\tilde\bX_i; \bdelta\rangle \cdot \mathbb{I}(|\langle\tilde\bX_i; \bdelta\rangle|\leq \mathfrak{c})$. By definition, $|\alpha_i|\leq \mathfrak{c}$. Now, for every realization of the random variables $\tilde\bX_1, \dots, \tilde\bX_n$ we can obtain
\[
\mathbb{E}_{\zeta}\Bigg[\sup_{\bdelta \in \mathcal{C}'(\tilde \theta, \tilde \eta,T)} \Big|\frac{1}{n}\sum_{i=1}^n \zeta_i f(\tilde \bX_i;\bdelta)\Big|\Bigg] \leq \mathfrak{c} \mathbb{E}_{\zeta}\Bigg[\sup_{\bdelta \in \mathcal{C}'(\tilde \theta, \tilde \eta,T)} \Big|\frac{1}{n}\sum_{i=1}^n \zeta_i \langle\tilde \bX_i;\bdelta\rangle\Big|\Bigg].
\]
Now, taking expectation of both sides with respect to $\tilde\bX_i$'s, using the tower property, and combining with (\ref{ineq1}) we obtain
\begin{align*}
  \mathbb{E}[\widetilde W_T] &\leq 8\mathfrak{c}\mathbb{E}\Bigg[\sup_{\bdelta \in \mathcal{C}'(\tilde \theta, \tilde \eta,T)} \Big|\frac{1}{n}\sum_{i=1}^n \zeta_i \langle\tilde \bX_i;\bdelta\rangle\Big|\Bigg]\\
  &\leq 8\mathfrak{c}\mathbb{E}\Bigg[\|\tilde\bSigma_{R}\|_{\op}\sup_{\bdelta\in \mathcal{C}'(\tilde\theta,\tilde\eta, T)}\|\bdelta\|_* \Bigg]\\
  &\leq 8\mathfrak{c}\sqrt{\tilde\eta}\mathbb{E}\Bigg[\|\tilde\bSigma_{R}\|_{\op}\sup_{\bdelta\in \mathcal{C}'(\tilde\theta,\tilde\eta, T)}\|\bdelta\|_F \Bigg]\\
    &\leq 8\mathfrak{c}\frac{\sqrt{l}}{d}\sqrt{\tilde\eta}\mathbb{E}\Bigg[\|\tilde\bSigma_{R}\|_{\op}\sup_{\bdelta\in \mathcal{C}'(\tilde\theta,\tilde\eta, T)}\|\sqrt{\bW}^{-1}\circ\bdelta\|_{L^2(\Pi)} \Bigg]\\
  &\leq 8\mathfrak{c}\frac{\sqrt{l}}{d}\sqrt{\tilde\eta\xi T}\mathbb{E}\Big[\|\tilde\bSigma\|_{\op} \Big].
\end{align*}

In the above, we used definition of $\mathcal{C}'(\tilde \theta, \tilde \eta,T)$ and also the following derivation:
\begin{align}
 \|\bdelta\|^2_F &= \|\sqrt{\bW}\circ \bM\|^2_F = \|\sqrt{\bP}\frac{\sqrt{\bW}}{\sqrt{\bP}}\circ \bM\|^2_F \nonumber \\ 
& \leq \frac{l}{d^2}\|\bM\|_{L^2(\Pi)}^2 = \frac{l}{d^2}\|\sqrt{\bW}^{-1}\circ\bdelta\|_{L^2(\Pi)}^2 \text{\hspace{5mm} for some $\bM$ such that $\bdelta = \sqrt{\bW}\circ \bM$}. \label{rmk1}
\end{align}
In the above, we have used Definition \ref{def_l} and the following property:
\begin{align*}
   \|\bA\|_{L^2(\Pi)}^2 &= \mathbb{E}[\langle \bX_i, \bA\rangle^2] = \mathbb{E}_{\Pi}[\mathbb{E}_{\xi}\langle \xi_i\cdot e_{r_i}e_{c_i}^T, \bA\rangle^2|r_i, c_i] = \mathbb{E}_{\Pi}[\mathbb{E}_{\xi}[\bA_{r_i, c_i}^2\xi_i^2] | r_i, c_i] = \mathbb{E}_{\Pi}[d^2\bA^2_{r_i, c_i}] \\
   &= d^2\sum\bA^2_{r_i, c_i}\bP_{r_i, c_i} = d^2\|\sqrt{\bP}\circ\bA\|_F^2.
\end{align*}
We can, now, use $2ab\leq a^2 + b^2$ to get
\[
\mathbb{E}[\widetilde W_T]\leq \frac{8}{9}(\frac{5\xi T }{24}) + 87\frac{l}{d^2}\tilde\eta\mathfrak{c}^2\mathbb{E}\Big[\|\tilde\bSigma_{R}\|_{\op} \Big]^2.
\]
Next, we turn to finding an upper bound for the variance of $\frac{1}{n}\Sigma_{i=1}^n f(\tilde\bX_i;\bdelta) - \mathbb{E}[f(\tilde\bX_i;\bdelta)]:$
\[
\Var(f(\tilde\bX_i;\bdelta) - \mathbb{E}[f(\tilde\bX_i;\bdelta)])\leq \mathbb{E}[f(\tilde\bX_i;\bdelta)^2]\leq \mathfrak{c}^2\cdot \mathbb{E}\big[\langle\tilde\bX_i, \bdelta\rangle^2\big] =  \mathfrak{c}^2\cdot\|\sqrt{\bW}^{-1}\circ\bdelta\|_{L^2(\Pi)}^2.
\]

 Therefore, we have that 
 \[
 \sup_{\bdelta\in \mathcal{C}'(\tilde\theta, \tilde\eta, T)}\frac{1}{n}\Var(f(\tilde\bX_i;\bdelta) - \mathbb{E}[f(\tilde\bX_i;\bdelta)])\leq \frac{\mathfrak{c}^2}{n}\cdot \sup_{\bdelta\in \mathcal{C}'(\tilde\theta,\tilde\eta,T)}\|\sqrt{\bW}^{-1}\circ\bdelta\|_{L^2(\Pi)}^2 \leq \frac{\xi T\mathfrak{c}^2}{n}.
 \]
 Finally, noting that $\frac{1}{n}f(\tilde\bX_i;\bdelta)\leq \frac{1}{n}\mathfrak{c}^2$ almost surely, we can use Massart's inequality to conclude that
 \begin{align*}
   \mathbb{P}(\widetilde W_T\geq \frac{5\xi T}{24} + \beta) &= \mathbb{P}(\widetilde W_T\geq \frac{5\xi T}{24} + 93\frac{l}{d^2}\eta \mathfrak{c}^2\mathbb{E}[\|\tilde\bSigma_R\|_{\op}]^2)\\
 &\leq \mathbb{P}\Bigg(\widetilde W_T\geq \frac{18}{17}\mathbb{E}[\widetilde W_T] + \frac{1}{17}\Big(\frac{5\xi T}{24}\Big)\Bigg)\leq \exp\Bigg(-\frac{Cn\xi T}{\mathfrak{c}^2}\Bigg),   
 \end{align*}
 for some numerical constant $C > 0$.
 \QED
\endproof

Lemma \ref{lemma1} entails that 
\[
\mathbb{P}(\mathcal{B}_k)\leq \exp\Bigg(-\frac{Cn\xi^k\tilde\theta}{\mathfrak{c}^2}\Bigg)\leq \exp\Bigg(-\frac{Cnk\log(\xi)\tilde\theta}{\mathfrak{c}^2}\Bigg).
\]
Therefore, by setting the numerical constant $C>0$ appropriately, the union bound implies that
\[
\mathbb{P}(\mathcal{B})\leq\sum_{k=1}^{\infty}\mathbb{P}(\mathcal{B}_k)\leq \sum_{k=1}^{\infty}\exp\Bigg(-\frac{Cnk\tilde\theta}{\mathfrak{c}^2}\Bigg) = \frac{\exp(-\frac{Cn\tilde\theta}{\mathfrak{c}^2})}{1 - \exp(-\frac{Cn\tilde\theta}{\mathfrak{c}^2})}.
\]
Finally, assuming that $Cn\tilde\theta>\mathfrak{c}^2$, we get that \[\mathbb{P}(\mathcal{B})\leq 2\exp\Bigg(-\frac{Cn\tilde\theta}{\mathfrak{c}^2}\Bigg),
\]
which completes the proof.
\QED
\endproof

Note that Lemma \ref{RSC} states RSC holds for $\mathcal{C'}(\tilde \theta, \tilde \eta)$ which is slightly different than the set $\mathcal{C}(\tilde \nu, \tilde \eta)$ defined earlier. But, using Assumption \ref{def_l}, we can see that
\[
\mathcal{C}(\tilde \nu, \tilde \eta)\subseteq \mathcal{C'}(\frac{\tilde\nu d^2}{l}, \tilde \eta).
\]

This can be shown as follows. Suppose $\bdelta \in \mathcal{C}(\tilde \nu, \tilde \eta)$, then $\|\bdelta\|_F^2 \ge \tilde\nu$. Inequality (\ref{rmk1}) implies $$\frac{l}{d^2}\|\sqrt{\bW}^{-1}\circ\bdelta\|_{L^2(\Pi)}^2 \geq \|\bdelta\|_F^2 \ge \tilde\nu,$$ so $\|\sqrt{\bW}^{-1}\circ\bdelta\|_{L^2(\Pi)}^2 \geq \frac{\tilde\nu d^2}{l}$, therefore, $\bdelta \in \mathcal{C'}(\frac{\tilde \nu d^2}{l}, \tilde \eta)$. As a result, Lemma \ref{lemma:RSC} follows.

\section{Bounding $\|\tilde\bSigma\|_{\op}$ and $\|\tilde\bSigma_R\|_{\op}$ with matrix Berstein inequality}
\label{append:bern}
The next proposition is a variant of the Bernstein inequality that will be used to show that Condition (\ref{lambdaass}) for $\tilde\lambda$ is guaranteed to hold with high probability as well.
\begin{proposition}
\label{matrixbernstein}
Let $(\bZ_i)_{i=1}^n$ be a sequence of $d\times d$ independent random matrices with zero mean, such that 
\[
\mathbb{E}\Big[\frac{\|\bZ_i\|_{\op}}{\delta}\Big] \leq e \hspace{1cm} \forall i \in [n],
\]
and 
\[
\sigma_{\bZ} = \max\Bigg\{\Big\|\frac{1}{n} \sum_{i=1}^n\mathbb{E}[\bZ_i\bZ_i^T]\Big\|_{\op}, \Big\|\frac{1}{n} \sum_{i=1}^n\mathbb{E}[\bZ_i^T\bZ_i]\Big\|_{\op}\Bigg\}^{\frac{1}{2}},
\]

for some positive values $\delta$ and $\sigma_z$. Then, there exits numerical constant $C>0$ such that, for all $t>0$
\[
\Bigg\|\frac{1}{n} \sum_{i=1}^n \bZ_i\Bigg\| \leq C\max\Bigg\{\sigma_{\bZ}\sqrt{\frac{t+\log(2d)}{n}}, \delta\big(\log \frac{\delta}{\sigma_{\bZ}}\big)\frac{t+\log(2d)}{n}\Bigg\},
\]
with probability at least $1 - \exp(-t)$.
\end{proposition}

Next, based on the above proposition, we adapted the following two lemmas from \cite{klopp}.

\begin{lemma}[Adapted from Lemma 5 of \cite{klopp}]
\label{corollaryextended1}  Let $(\bZ)_{i = 1}^n$ be a sequence of $d\times d$ independent random matrices with zero mean. Suppose that $a\leq\sigma_z\leq b$ and $\mathbb{E}\Big[\exp\Big(\frac{\|\bZ_i\|_{\op}}{\delta}\Big)\Big]\leq e$ for every $i\in [n]$. Then
\begin{align}
\label{matrixbernsteincor}
 \mathbb{P}\Bigg(\lambda\leq \|\frac{1}{n}\sum_{i=1}^n \bZ_i\|_{\op}\Bigg) &\leq 2d\exp\Bigg[-C\lambda n\Bigg(\frac{\lambda}{b^2} \wedge \frac{1}{\delta(\log\frac{\delta}{a})}\Bigg)\Bigg].
\end{align}
\end{lemma}
Let $\bZ_i = \epsilon_i\tilde\bX_i$, then the above lemma can be applied to $\tilde\bSigma =\frac{1}{n}\sum_{i=1}^n \bZ_i$.

\proof{Proof of Lemma \ref{corollaryextended1}.}
We can apply Proposition \ref{matrixbernstein} to get: with probability at least $1 -\exp(-t)$,
\begin{align*}
  \|\frac{1}{n}\sum_{i=1}^n \bZ_i\|_{\op} &\leq C\max\Bigg\{\sigma_z\sqrt{\frac{t + log(2d)}{n}}, \delta(\log\frac{\delta}{\sigma_z}) \frac{t+\log(2d)}{n}\Bigg\} \\
  &\leq C \max\Bigg\{ \underbrace{b\sqrt{\frac{t + log(2d)}{n}}}_{\circled{1}}, \underbrace{\delta(\log\frac{\delta}{a}) \frac{t+\log(2d)}{n}}_{\circled{2}}\Bigg\}
\end{align*}

We can find $t = t^*$ such that $\circled{1} = \circled{2}$. $\circled{1} = \circled{2}$ implies that
\[
    b\sqrt{\frac{t+\log(2d)}{n}} = \delta(\log\frac{\delta}{a})\frac{t+\log(2d)}{n}
\]
So $$t^* =-\log(2d)+ \frac{b^2n}{\delta^2(\log\frac{\delta}{a})^2}.$$
\textbf{When $t\leq t^*$}, we have $\circled{1}\leq \circled{2}$. 
Then, with probability at most $\exp(-t)$,
\[
 \|\frac{1}{n}\sum_{i=1}^n \bZ_i\|_{\op} \geq Cb\sqrt{\frac{t+\log(2d)}{n}}.
\]
Let $\lambda = Cb\sqrt{\frac{t+\log{2d}}{n}}$, then $$t = -\log(2d) + \frac{\lambda^2n}{C^2 b^2}\text{ \hspace{0.5cm} and \hspace{0.5cm}} \exp(-t) = 2d\exp(-\frac{\lambda^2n}{C^2b^2}).$$

So \[
\mathbb{P}\Bigg(  \|\frac{1}{n}\sum_{i=1}^n \bZ_i\|_{\op} \geq \lambda \Bigg)\leq 2d \exp(-\frac{\lambda^2n}{C^2b^2}).
\]

\textbf{When $t\geq t^*$}, we have $\circled{2}\geq \circled{1}$. 

Then, with probability at most $\exp(-t)$,
\[
 \|\frac{1}{n}\sum_{i=1}^n \bZ_i\|_{\op} \geq C\delta(\log\frac{\delta}{a})\frac{t+\log(2d)}{n}.
\]
Let $\lambda = C\delta(\log\frac{\delta}{a})\frac{t+\log(2d)}{n}$, then $$t = -\log(2d) + \frac{\lambda n}{C\delta(\log\frac{\delta}{a})}\text{ \hspace{0.5cm} and \hspace{0.5cm}} \exp(-t) = 2d\exp(-\frac{\lambda n}{C \delta(\log\frac{\delta}{a})}).$$

So $$\mathbb{P}\Bigg(  \|\frac{1}{n}\sum_{i=1}^n \bZ_i\|_{\op} \geq \lambda\Bigg)\leq 2d \exp(-\frac{\lambda n}{C\delta(\log\frac{\delta}{a})}).$$

It follows that
\begin{align*}
 \mathbb{P}\Bigg(\|\frac{1}{n}\sum_{i=1}^n \bZ_i\|_{\op}\geq \lambda\Bigg) &\leq \Big[2d\exp\big(-\frac{\lambda^2 n}{C^2b^2}\big)\Big]\vee \Big[2d\exp\big(-\frac{\lambda n}{C\delta(\log\frac{\delta}{a})}\big)\Big] \\
 &= 2d\exp\Bigg[-C\lambda n\Bigg(\frac{\lambda}{b^2} \wedge \frac{1}{\delta(\log\frac{\delta}{a})}\Bigg)\Bigg].
\end{align*}
\QED
\endproof

\begin{lemma}[Adapted from Lemma 6 of \cite{klopp}]
\label{corollaryextended2}
If Condition (\ref{matrixbernsteincor}) holds and $n\geq \frac{\delta^2(\log\frac{\delta}{a})^2\log2d)}{b^2}$, then
\[
\mathbb{E}\|\frac{1}{n} \sum_{i=1}^n \bZ_i\|_{\op} \leq C \sqrt{\frac{2e\log(2d)b^2}{n}}.
\]
\end{lemma}

\proof{Proof of Lemma \ref{corollaryextended2}.}

Set $$\nu_1 = \frac{n}{C^2b^2} \text{\hspace{0.5cm}  and  \hspace{0.5cm}}\nu_2 = \frac{n}{C\delta(\log\frac{\delta}{a})}.$$

By H\"{o}lder's inequality, we get
\[
\mathbb{E}\|\frac{1}{n}\sum_{i=1}^n\zeta_i \bX_i\|\leq \Bigg(\mathbb{E}\|\frac{1}{n}\sum_{i=1}^n\zeta_i \bX_i\|^{2\log (d)}\Bigg)^{1/(2\log(d))}.
\]

The inequality (\ref{matrixbernsteincor}) imply that 
\begin{align}
\label{integral}
 & \Bigg(\mathbb{E}\|\frac{1}{n}\sum_{i=1}^n\zeta_i\bX_i\|^{2\log (d)}\Bigg)^{1/(2\log(d))} \nonumber\\
=& \Bigg(\int_0^{+\infty}\mathbb{P}\Bigg(\|\frac{1}{n}\sum_{i=1}^n\zeta_i \bX_i\|>\lambda^{1/(2\log(d))}\Bigg)d\lambda\Bigg)^{1/(2\log(d))}\nonumber\\
\leq &\Bigg(d\int_0^{+\infty} \exp\{-t^{1/\log(d)}\nu_1\}d\lambda + d\int_0^{+\infty}\exp\{-\lambda^{1/(2\log(d))}d\lambda\}d\lambda\Bigg)^{1/(2\log(d))}\nonumber\\
\leq & \sqrt{e}(\log(d)\nu_1^{-\log(d)}\Gamma(\log(d)) + 2\log(d)\nu_2^{-2\log(d)}\Gamma(2\log(d)))^{1/(2\log(d))}.
\end{align}

The Gamma-function satisfies the following bound:
for $x\geq 2$, $\Gamma(x)\leq (\frac{x}{2})^{x-1}$. Plugging this into (\ref{integral}), we compute 
\begin{align*}
 \mathbb{E}\|\frac{1}{n}\sum_{i=1}^n\zeta_i \bX_i\|\leq 
\sqrt{e}((\log(d))^{\log(d)}\nu_1^{-\log(d)}2^{1-\log(d)} + 2(\log(d))^{2\log(d)}\nu_2^{-2\log(d)})^{1/(2\log(d))}.
\end{align*}

Then $$n\geq \frac{\delta^2(\log\frac{\delta}{a})^2\log(d)}{b^2}$$ implies that $\nu_1\log(d)\leq \nu_2^2$, which then entails
\[
\mathbb{E}\|\frac{1}{n} \sum_{i=1}^n \bX_i\|_{\op} \leq C \sqrt{\frac{2e\log(d)b^2}{n}}.
\]
\QED
\endproof

The next step is to use Proposition \ref{matrixbernstein} for $\bZ_i:=\epsilon_i\tilde\bX_i$ to find a tail bound inequality for $\mathbb{P}(\lambda<3\|\bSigma\|_{\op})$. Define $\delta = \frac{d\sigma e}{\sqrt{p_{\min}/l}(e-1)}$ and let $G_1$ and $G_2$ be two independent standard normal random variables. Then, it follows that
\begin{align*}
    \mathbb{E}\Big[\exp\Big(\frac{\|\bZ_i\|_{\op}}{\delta}\Big)\Big] &= \mathbb{E}\Big[\Big(\frac{d\sigma|G_1G_2|}{\delta\sqrt{\bW_{r_i c_i}}}\Big)\big| r_i, c_i\Big] \leq \mathbb{E}\Big[\Big(\frac{d\sigma(G_1^2 + G_2^2)}{2\delta\sqrt{\bW_{r_i c_i}}}\Big)\big| r_i, c_i\Big]\\
     &= \mathbb{E}\Big[\Big(\frac{d\sigma|G_1^2|}{2\delta\sqrt{\bW_{r_i c_i}}}\Big)\big| r_i, c_i\Big]^2 \leq \mathbb{E}\Big[\Big(\frac{d\sigma G_1^2}{2\delta\sqrt{p_{\min}/l}}\Big)\big| r_i, c_i\Big]^2\\
        &\leq \mathbb{E}\Big[\Big(\frac{(e-1)G_1^2}{2e}\Big)\big| r_i, c_i\Big]^2 =\Bigg[\frac{1}{\sqrt{1 - \frac{e-1}{e}}}\Bigg]^2 = e.
\end{align*}

Next, we need to figure out $\mathbb{E}[\bZ_i\bZ_i^{\top}]$ and $\mathbb{E}[\bZ_i^{\top}\bZ_i]$. 
\begin{align*}
    \mathbb{E}[\bZ_i\bZ_i^{\top}] &= \mathbb{E}[\epsilon_i\tilde\bX_i\epsilon_i\tilde\bX_i^{\top}] = \mathbb{E}[\epsilon_i^2\tilde\bX_i\tilde\bX_i^{\top}] = \mathbb{E}[\epsilon_i^2\xi_i^2(\sqrt{\bW}^{-1}\circ e_{r_i}e_{c_i}^{\top})(\sqrt{\bW}^{-1}\circ e_{r_i}e_{c_i}^{\top})^{\top}| r_i, c_i]\\
    &=\mathbb{E}[\epsilon_i^2\xi_i^2(\sqrt{\bW}^{-1}_{r_i c_i}e_{r_i}e_{c_i}^{\top})(\sqrt{\bW}^{-1}_{r_i c_i}e_{c_i}e_{r_i}^{\top})| r_i, c_i] =\sigma^2 d^2\mathbb{E}[\bW^{-1}_{r_i c_i}e_{r_i}e_{r_i}^{\top}| r_i, c_i]\\
    &=\sigma^2 d^2\sum_j \sum_k \bP_{jk} \bW^{-1}_{jk}e_j e_j^{\top} = \sigma^2 d^2\text{Diag} \Bigg[\sum_{k=1}^d\frac{\bP_{1k}}{\bW_{1k}} \hspace{0.5cm} \sum_{k=1}^d\frac{\bP_{2k}}{\bW_{2k}} \hspace{0.5cm} \dots \hspace{0.5cm} \sum_{k=1}^d\frac{\bP_{dk}}{\bW_{dk}}\Bigg]\\
    &= \sigma^2 d^2\text{Diag rowsum}\Bigg(\frac{\bP}{\bW}\Bigg)
\end{align*}
Similarly, 
\begin{align*}
    \mathbb{E}[\bZ_i^{\top}\bZ_i] &= \mathbb{E}[\epsilon_i\tilde\bX_i^{\top}\epsilon_i\tilde\bX_i] = \mathbb{E}[\epsilon_i^2\tilde\bX_i^{\top}\tilde\bX_i] = \mathbb{E}[\epsilon_i^2\xi_i^2(\sqrt{\bW}^{-1}\circ e_{r_i}e_{c_i}^{\top})^{\top}(\sqrt{\bW}^{-1}\circ e_{r_i}e_{c_i}^{\top})| r_i, c_i]\\
    &=\mathbb{E}[\epsilon_i^2\xi_i^2(\sqrt{\bW}^{-1}_{r_i c_i}e_{c_i}e_{r_i}^T)(\sqrt{\bW}^{-1}_{r_i c_i}e_{r_i}e_{c_i}^{\top})| r_i, c_i] =\sigma^2 d^2\mathbb{E}[\bW^{-1}_{r_i c_i}e_{c_i}e_{c_i}^{\top}| r_i, c_i]\\
    &=\sigma^2 d^2\sum_j \sum_k \bP_{jk} \bW^{-1}_{jk}e_k e_k^{\top} = \sigma^2 d^2\text{Diag} \Bigg[\sum_{j=1}^d\frac{\bP_{j1}}{\bW_{j1}} \hspace{0.5cm} \sum_{j=1}^d\frac{\bP_{j2}}{\bW_{j2}} \hspace{0.5cm} \dots \hspace{0.5cm} \sum_{j=1}^d\frac{\bP_{jd}}{\bW_{jd}}\Bigg]\\
    &= \sigma^2 d^2\text{Diag columnsum}\Bigg(\frac{\bP}{\bW}\Bigg)
\end{align*}    

So 
\begin{align*}
   \sigma_z:&= \max\Bigg\{\|\mathbb{E}[\bZ_i\bZ_i^{\top}]\|_{\op},\|\mathbb{E}[\bZ_i^{\top}\bZ_i]\|_{\op}\Bigg\}^{1/2} \\
   &= \sigma d\max\Bigg\{ \sum_{k=1}^d\frac{\bP_{1k}}{\bW_{1k}}, \sum_{k=1}^d\frac{\bP_{2k}}{\bW_{2k}}, \dots, \sum_{k=1}^d\frac{\bP_{dk}}{\bW_{dk}}, \sum_{j=1}^d\frac{\bP_{j1}}{\bW_{j1}}, \sum_{j=1}^d\frac{\bP_{j2}}{\bW_{j2}},\dots, \sum_{j=1}^d\frac{\bP_{jd}}{\bW_{jd}} \Bigg\} ^{1/2} \\
   &= \sigma d\max\Bigg\{\text{rowsum}\Bigg(\frac{\bP}{\bW}\Bigg), \text{colsum}\Bigg(\frac{\bP}{\bW}\Bigg)\Bigg\} ^{1/2}.
\end{align*}

Note that, 
\[
\sigma_z^2\leq \sigma^2 d^3l\quad\text{ and } \quad \sigma_z^2\geq \frac{\sigma^2d^3}{l}.
\]

So, in the case $$\bZ_i =\epsilon_i\tilde\bX_i \text{ \hspace{0.5cm} and \hspace{0.5cm}} a = \sqrt{\frac{\sigma^2d^3}{l}}\text{  \hspace{0.5cm} and \hspace{0.5cm}} b = \sqrt{\sigma^2 d^3l} \text{  \hspace{0.5cm} and \hspace{0.5cm}} \delta = \frac{ed\sigma}{(e-1)\sqrt{p_{\min}/l}},$$

we have the following by applying Lemma \ref{corollaryextended1}:
\begin{align}
    \mathbb{P}\Bigg(\|\tilde\bSigma\|_{\op}\geq \lambda\Bigg) &\leq  2d\exp\Bigg[-C n\lambda \Bigg(\frac{\lambda}{\sigma^2 d^3 l} \wedge \frac{1}{\frac{d\sigma\sqrt{l}}{\sqrt{p_{\min}}} \frac{e}{e-1}(\log\frac{ed\sigma \sqrt{l}\sqrt{l}}{(e-1)\sqrt{p_{\min}\sigma^2 d^3}})}\Bigg)\Bigg] \nonumber\\
    &\leq 2d\exp\Bigg[-\frac{C n\lambda }{d\sigma}\Bigg( \frac{\lambda}{\sigma d^2 l}\wedge \frac{\sqrt{p_{\min}}}{\sqrt{l}\log (\frac{l^2}{p_{\min}d})}\Bigg)\Bigg]. \label{Sigma}
\end{align}

We can follow the same argument for $\bZ_i = \zeta_i \bX_i$, where $\delta = \frac{d e}{\sqrt{p_{\min}/l}(e-1)}$, and 

\begin{align*}
   \sigma_z:&= \max\Bigg\{\|\mathbb{E}[\bZ_i\bZ_i^{\top}]\|_{\op},\|\mathbb{E}[\bZ_i^{\top}\bZ_i]\|_{\op}\Bigg\}^{1/2} = d\max\Bigg\{\text{rowsum}\Bigg(\frac{\bP}{\bW}\Bigg), \text{colsum}\Bigg(\frac{\bP}{\bW}\Bigg)\Bigg\} ^{1/2}.
\end{align*}

So 
\[
\sqrt{\frac{d^3}{l}} \leq \sigma_z\leq \sqrt{d^3l}.
\]

 \begin{align*}\frac{\delta^2(\log\frac{\delta}{a})^2\log(2d)}{b^2} &= \frac{d^2l}{p_{\min}}(\frac{e}{e-1})^2\Bigg(\log \frac{e}{e-1}\sqrt{\frac{d^2l}{(p_{\min}/l)d^3}}\Bigg)^2\log(2d)\frac{1}{d^3l}\\
 &\leq C\Bigg(\log \frac{l^2}{p_{\min}d} \Bigg)^2\frac{\log(2d)}{p_{\min}d}.
 \end{align*}
 Plugging in $b$, we have
 \begin{align*}
     C\sqrt{\frac{2e\log(d)b^2}{n}} &\leq C'\sqrt{\frac{\log(d)d^3l}{n}}.
 \end{align*}
Further, we have
$$n\geq C\Bigg(\log \frac{l^2}{p_{\min}d} \Bigg)^2\frac{\log(d)}{p_{\min}d}\geq \frac{C}{2}\Bigg(\log \frac{l^2}{p_{\min}d} \Bigg)^2\frac{\log(2d)}{p_{\min}d},$$
note that  we are implicitly using $\log(2d)\leq \log d^2 = 2\log d$ when $d\geq 2$.

We have 
\begin{align}
\label{SigmaR}
   \mathbb{E}\|\tilde\bSigma_R\|_{\op}\leq C'\sqrt{\frac{\log(d)d^3l}{n}}. 
\end{align}

\section{Proof of Theorem \ref{matcompthm}}
\label{appendix:thm1proof}
Following the recipe provided in \cite{nima}, we are left to demonstrate how to choose a norm $\mathfrak{N}(\cdot)$, which is guided by exponential Orlicz norm and obtain the constant $n^*$ and choose an appropriate $\mathfrak{c}$ such that Assumption \ref{assump_on_frak_c} holds.

For simplicity, we assume that $\epsilon_i\sim \mathcal{N}(0,\sigma^2)$ for all $i\in [n]$. Let $\bX_i = \xi_i\cdot e_{r_i}e_{c_i}^T$ where, for each $i$, $\xi_i$ is an independent $4d^2$-sub-Gaussian random variable that is also independent of $r_j$ and $c_j, j\in [n]$. If we set $\xi_i:=d$ almost surely, then $\|\xi_i\|_{\psi_2} = d/\sqrt{\log2} \leq 2d$, and so, satisfies our requirement. This corresponds to the problem studied in \cite{sahand}, which notes that $\xi_i$ has no statistical effect. Further, this way, we have $\mathbb{E}[\langle \bX_i, \bB\rangle] = 0$. Here we show the bounds for the slightly more general case of $\xi_i\sim \mathcal{N}(0, d^2)$.

In order to find a suitable norm $\mathfrak{N}(\cdot)$, we next study $\|\bN\|_{\psi_2(\Pi)}$ to see how heavy-tailed $\langle\bN, \bX_i\rangle$ is, where $\bN = \sqrt{\bW}\circ \bB$ for a given $\bB$. We have
\begin{align*}
   &\mathbb{E}\Bigg[\exp\Bigg(\frac{|\langle \bN, \bX_i\rangle|^2}{4d^4\|\bN\|^2_{\infty}}\Bigg)\Bigg] = \mathbb{E}_{\Pi}\Bigg[\mathbb{E}_{\xi} \Bigg[\exp\Bigg( \frac{\xi_i^2 \bN_{jk}^2}{4d^4\|\bN\|_{\infty}^2}\Bigg)\Bigg] \Big| j,k \Bigg] \\
   =& \mathbb{E}_{\Pi}\Bigg[\frac{1}{\sqrt{\Bigg(1-\frac{\bN_{jk}^2}{2d^2\|\bN\|_{\infty}^2}}\Bigg)_+} \Big |j, k\Bigg] \leq \mathbb{E}_{\Pi}\Bigg[\frac{1}{\sqrt{\Bigg(1-\frac{1}{2d^2}}\Bigg)_+} \Big |j, k\Bigg]\\
   \leq & \sum \bP_{jk}\sqrt{2}\leq \sqrt{2}\leq 2.
\end{align*}
Note that we have used Lemma \ref{A1} and the following inequality:
\[
\frac{\bN_{jk}^2}{2d^2\|\bN\|_{\infty}^2}\leq \frac{1}{2}.
\]

Therefore, 
\[
\|\bN\|_{\psi_2(\Pi)}\leq 2d^2\|\bN\|_{\infty},
\]
which guides selection of $\mathfrak{N}(\bN) = d^2\|\bN\|_{\infty}$ for any $\bB$ and $\bN =\sqrt{\bW}\circ\bB$ and $n^* = 2d^2\|\sqrt{\bW}\circ\bB^*\|_{\infty}$. We can now see that $\mathfrak{c} = \frac{9}{d}\sqrt{\frac{l}{p_{\min}}}$ fulfills Assumption \ref{assump_on_frak_c}. The reason is, given $ \mathfrak{N}(\bN) = d^2\|\sqrt{\bW}\circ \bB\|_{\infty} \leq 1$, we can condition on $r_i$ and $c_i$ and use Corollary \ref{corollary:constant} to obtain 
\begin{align*}
c_{d,2} &\leq 5\|\xi_i \bB_{r_i,c_i}\|_{\psi_2} = 5|\bB_{r_i,c_i}|\|\xi_i\|_{\psi_2} = 5\sqrt{\frac{8}{3}} d|\bB_{r_i,c_i}|=5\sqrt{\frac{8}{3}}d^2 |\bB_{r_i,c_i}|\sqrt{\frac{p_{\min}}{l}}\sqrt{\frac{l}{d^2p_{\min}}} \\ 
&\leq 5\sqrt{\frac{8}{3}}d^2 |\bB_{r_i,c_i}|\sqrt{\bW_{r_i,c_i}}\sqrt{\frac{l}{d^2p_{\min}}} \leq 5\sqrt{\frac{8}{3}}d^2\|\sqrt{\bW}\circ \bB\|_{\infty}\sqrt{\frac{l}{d^2p_{\min}}}\\
&=5\sqrt{\frac{8}{3}}\sqrt{\frac{l}{d^2p_{\min}}}\leq \frac{9}{d}\sqrt{\frac{l}{p_{\min}}} =\mathfrak{c}.
\end{align*}
So we can obtain 
\[
\mathbb{E}_{\xi}\Bigg[\xi_i^2\bB_{r_i, c_i}^2\cdot \mathbb{I}(|\xi_i \bB_{r_i, c_i}|\leq \frac{9}{d}\sqrt{\frac{l}{p_{\min}}}) \Big| r_i, c_i\Bigg]\geq \mathbb{E}_{\xi}\Bigg[\xi_i^2\bB_{r_i, c_i}^2\cdot \mathbb{I}(|\xi_i \bB_{r_i, c_i}|\leq c_{d,2}) \Big| r_i, c_i\Bigg]\geq \frac{\mathbb{E}[\xi_i^2\bB_{r_i, c_i}^2|r_i, c_i]}{2}
\]
Now we can take the expectation with respect to $r_i, c_i$ and use the tower property to show
\[
\mathbb{E}\Bigg[\langle \bB, \bX_i\rangle^2\cdot \mathbb{I}\Bigg(|\langle \bB, \bX_i\rangle|\leq \frac{9}{d}\sqrt{\frac{l}{p_{\min}}}\Bigg)\Bigg] \geq \frac{1}{2}\mathbb{E}[\langle\bB, \bX_i\rangle^2],
\]
which is equivalent to
\[
\mathbb{E}\Bigg[\langle\tilde \bX, \bN\rangle^2 \cdot \mathbb{I}\Bigg(\|\langle \tilde \bX, \bN\rangle\| \leq \frac{9}{d}\sqrt{\frac{l}{p_{\min}}}\Bigg)\Bigg]\geq \frac{1}{2}\mathbb{E}[\langle\tilde \bX, \bN\rangle^2].
\]

We can now combine Condition (\ref{Sigma}), (\ref{SigmaR}) and (\ref{beforeBernstein}) to obtain the following result:

for any $\lambda\geq C_3n^*\sqrt{\log(d)dl/n}$ and $n\geq C_3(\log\frac{l^2}{p_{\min}d})^2\frac{\log d}{p_{\min}d}$, the inequality 
\[
\|\mathbf{\widehat B} - \mathbf{B}^*\|_{L^2(\Pi)}^2\leq C_4\lambda^2\tilde rl^3,
\]
holds with probability at least 
\[
1-2d\exp\Bigg[-\frac{C n\lambda }{d\sigma}\Bigg( \frac{\lambda}{\sigma l}\wedge \frac{1}{\sqrt{l}\log (\frac{l^2}{d})}\Bigg)\Bigg] - \exp\Bigg(-\frac{C_5d^2n\lambda^2\tilde r}{{n^*}^2}\Bigg).
\]

In particular, setting 
\[
\lambda = C_6(\sigma \vee n^*)\sqrt{\frac{\rho l d}{n p_{\min}}}
\]
for some $\rho\geq \log d$. We have that
\[
\|\widehat\bB-\bB^*\|^2_{L^2(\Pi)}\leq C_1 (\sigma^2\vee {n^*}^2)\frac{d\rho l^4\tilde r}{n p_{\min}},
\]
with probability at least $1-\exp(-C_2\rho)$ for some $\rho\geq \log d$ and constants $C_1, C_2$ and $C_3$, whenever $n\geq C_3(\log\frac{l^2}{p_{\min}d})^2\frac{\log d}{p_{\min}d}$.

\section{Auxiliary proofs}
\label{append:aux}
\begin{lemma}
\label{A1}
Let $Z$ be a $\mathcal{N}(0, \sigma^2)$ random variable. Then, for all $\eta>0,$
\[
\mathbb{E}[e^{\eta Z^2}] = \frac{1}{\sqrt{(1-2\sigma^2\eta)_+}}.
\]
\end{lemma}

\begin{lemma}
Let $\bZ$ be a non-negative random variable such that $\|\bZ\|_{\psi_p} = \nu$ holds for some $p\geq 1$, and asuume $c>0$ is given. Then we have
\[
\mathbb{E}[Z^2\cdot \mathbb{I}(Z\geq c)] \leq (2c^2 + 4c\nu + 4\nu^2)\cdot \exp (-\frac{c^p}{\nu^p}).
\]
\end{lemma}

\begin{corollary}
Let $\bZ$ be a random variable satisfying $\|\bZ\|_{\psi_p} = \nu$ holds for some $p\geq 1$ and $\mathbb{E}[Z^2] = \sigma^2$. Then, for 
\begin{equation}
\label{constant}
   c_{\sigma, p}:=\nu\cdot \max\Bigg\{5, [10\log(\frac{2\nu^2}{\sigma^2})]^{\frac{1}{p}}\Bigg\},
\end{equation}

we have
\[\mathbb{E}[Z^2\cdot \mathbb{I}(|Z|\leq c_{\sigma, p})]\geq \frac{\mathbb{E}[Z^2]}{2}.
\]
\end{corollary}

\begin{corollary}
\label{corollary:constant}
Let $Z$ be a $\mathcal{N}(0, \sigma^2)$ random variable. Then, the constant $c_{\sigma, 2}$ defined in (\ref{constant}) satisfies $c_{\sigma, p}\leq 5\|Z\|_{\psi_2}.$
\end{corollary}

The following lemma gives rise to the proximal mapping we came up earlier. 

\begin{lemma}
 \label{thm:1}
  $$\mathcal{S}_{\lambda}(\bM) = \argmin_{\bM}\frac{1}{2} \|\bM-\bN\|^2_F + \lambda \|\bM\|_*.$$
\end{lemma}

\proof {Proof. (This proof is adapted from Cai et al. (2008))}
Since the function $$h_0(\bM):= \frac{1}{2} \|\bM-\bN\|^2_F + \lambda \|\bM\|_*.$$ is strictly convex, it is easy to see that there exists a unique minimizer, call it $\widehat \bM$. Now $\widehat \bM$ minimizes $f(\bM)$ if and only if 0 is a subgradient of the functional $f$ at the point $\widehat \bM$, i.e., $$\mathbf{0}\in \partial f(\widehat \bM).$$
We have
\[\partial f(\widehat \bM) = \lambda \partial \|\widehat \bM\|_* + \widehat \bM -\bN,
\]
where $\partial \|\widehat \bM\|_*$ is the set of subgradients of the trace norm.

It is known that 

$$\partial \| \bM\|_* = \{ \bU\bV^* + \bW: \bW\in \mathbb{R}^{n \times m}, \bU^*\bW = 0, \bW\bV = 0, \|\bW\|_{op} \leq 1\},$$

where $\bM = \bU\bSigma \bV^*$ is the SVD of $\bM$.

Now we want to show $\widehat \bM = \mathcal{S}_{\lambda}(\bN)$. We first decompose the SVD of $\bN$ as:
\[
\bN = \bU_0\bSigma_0\bV_0^* + \bU_1\bSigma_1\bV_0^*,
\]
where $\bU_0, \bV_0 (resp. \bU_1, \bV_1)$ are the singular vectors associated with singular values greater than $\lambda$ (resp. smaller than or equal to $\lambda$).

With these notation, we have $\mathcal{S}_{\lambda}(\bN) = \bU_0(\bSigma_0 - \lambda \bI) \bV_0^*$, and therefore, $\bN - \mathcal{S}_{\lambda}(\bN) = \lambda(\bU_0 \bV_0^* +\bW)$, where $\bW = \lambda^{-1} \bU_1 \bSigma_1 \bV_1^*$.

By definition, $\bU_0^* \bW = 0, \bW\bV_0 = 0$ and since the diagonal elements of $\bSigma_1$ have magnitudes bounded by $\lambda$, we also have $\|\bW\|_2 \leq 1$. Hence, $\bN - \mathcal{S}_{\lambda}(\bN)\in \lambda\partial \|\widehat \bM\|_* $, which concludes the proof.
\QED
\endproof

\end{APPENDICES}

\end{document}